\pdfoutput=1
\documentclass[1pt]{iopart}

\usepackage[utf8]{inputenc}
\usepackage[T1]{fontenc}
\usepackage{graphicx}
\usepackage{subfigure}
\usepackage{fullpage}
\usepackage{hyperref}
\usepackage{xspace}
\usepackage{microtype}
\usepackage{soul}

\usepackage{algorithmic}
\usepackage{algorithm}
\newcommand{\procName}{Procedure}
\floatname{algorithm}{\procName}

\providecommand{\eg}{\emph{e.g.}\xspace}
\providecommand{\etal}{\emph{et al.}\xspace}

\newcommand{\golRule}[1]{\ensuremath{\mathcal{R}_{#1}}}

\begin{document}
	

\title{Rule switching mechanisms in the Game of Life with synchronous and asynchronous updating policy}
\author{Jaros{\l}aw Adam Miszczak}%
\ead{jmiszczak@iitis.pl}
\address{{Institute of Theoretical and Applied Informatics, Polish Academy
		of Sciences}, {Baltycka 5}, {44-100}, {Gliwice}, {Poland}}%

\begin{abstract}
The emergence of complex structures in the systems governed by a simple set of rules is among the most fascinating aspects of Nature. The particularly powerful and versatile model suitable for investigating this phenomenon is provided by cellular automata, with the Game of Life being one of the most prominent examples. However, this simplified model can be too limiting in providing a tool for modelling real systems. To address this, we introduce and study an extended version of the Game of Life, with the dynamical process governing the rule selection at each step. We show that the introduced modification significantly alters the behaviour of the game. We also demonstrate that the choice of the synchronization policy can be used to control the trade-off between the stability and the growth in the system.
\end{abstract}

\section{Introduction}

Cellular automata (CA) provide a powerful model of computation, which was developed to mimic the emergence of complex phenomena, including self-replication~\cite{vonneumann1966theory}, in the systems governed by a finite set of simple rules. CA have found applications in many branches of science~\cite{schiff2011cellular}, including physics, where they were utilized as a model for thermodynamic systems~\cite{chopard1998cellular} and to develop a formulation of quantum mechanics~\cite{hooft2016cellular}, in computer science, where this model has been used to study various aspects of computation~\cite{wolfram1994cellulara,wolfram1984cellular,rendell2011universal}, or even in chemistry as a model for complex reactions~\cite{kier1999cellular,korte2013cellular}. Depending on the characteristic of the studied phenomena, the variety of modifications of the base model has been proposed~\cite{pca2018}, and the recent survey of different approaches to cellular automata is provided by Bhattacharjee \etal~\cite{bhattacharjee2018survey}.

Significant research effort has been devoted to investigate the significance of
the updating policy for the systems modelled by CA. The study of asynchronous cellular automata has been actively developed, and the summary of recent developments is provided in a review paper by Fat\`es \cite{fates2014guided}. In particular, the
effect of the update policy on the stability of cellular automata has been
investigated by Baetens \etal~\cite{baetens2012effect}. The detailed comparison of synchronous,
asynchronous, and sequential updating for a class of cellular automata was
provided by Reia and Kinouchi  \cite{reia2015nonsynchronous}, demonstrating that the changes in the update policy destroy the typical structures in the studied automata. Similarly,  the stochastic update policy --
including the random update or disruption of the cell-to-cell transmission of
information --  has been used to probe the robustness of the behaviour of
elementary cellular automata \cite{boure2012probing}. As a result, a wide variety of results has been observed for the case of 1D cellular automata. The detailed study of the asynchronous
behaviour, based on the disruption of information, is presented by Bour{\'{e}} \etal\cite{boure2011robustness}. However, in that case the main focus was on random disruptions of  the transmission of information between cells.

Among CA, the prominent example is provided by the Game of Life (GoL), which is an example of 2D CA capable of universal computation~\cite{berlekamp2001winning}. 
As the synchronous update is not suited to provide a model of many realistic systems, alternative models for obtaining behaviour observer in the GoL have been proposed~\cite{peper2010variations}. In \cite{blok1999synchronous} a modification of the GoL has been introduced in which
only a fraction of patches is updated during each time step. The updating scheme proposed in \cite{blok1999synchronous} has been utilized in the asynchronous GoL described in \cite{lee2004asynchronous}, where  the statistical properties of the introduced model have been investigated.

Concerning the motivation behind the chosen updating policy, it can be
influenced by the particular research topic. For example, in
\cite{poindron2021general} the asynchronous updating was based on the mutual
influence of agents. Probabilistic cellular automata were investigated using motivation from physics, computer science, and biology~\cite{billings2003indentification,agapie2014probabilisti,mairesse2014probabilistic}. 
More recently, a systematic study of cellular automata with random rules has been provided \cite{gravner2021periodic,gravner2022one}. In particular, Aguilera-Venegas \etal \cite{aguileravenegas2019probabilistic}  considered a probabilistic version of the GoL. This model has also been exploited to study the power of quantum annealing devices~\cite{gabor2021probabilistic}. 

Several variants of the GoL were studied to explore the possibility of extending this model. For example, a version including asymmetry in the interaction has been proposed  \cite{mullick2019effect}, where a system in which agents are biased towards their nearest neighbours and annihilate as they meet was proposed.  Moreover, a self-referencing variant has been studied to model the traces each generation leaves in the environment \cite{pavlic2014self}. Such traces affect the shape of the next generations of the population. This modification overcomes the lack of state feedback, which is one of the limitations of conventional cellular automata, but can be observed in natural evolving systems.

Another opportunity to enrich the GoL is to extended the space of possible moves. In particular, the  continuous variant of the GoL was proposed \cite{chan2019lenia,chan2020lenia} and it was observed that this system supports a great diversity of complex structures. As one of the emerging paradigms of computing is based on the quantum mechanics~\cite{miszczak2012high-level}, the quantum version of the GoL has also been introduced~\cite{arrighi2010quantum,bleh2012quantum}, and studied in the context of non-classical correlations~\cite{ney2022entanglement}. Moreover, it it has been suggested that the GoL can be realized on quantum computing platforms~\cite{hillberry2021entangled} and result in behaviour leading to the formation of small-world mutual information networks~\cite{jones2022small}.

In this work, we follow this track of modifications, and we study the model of 2D cellular automata enriched by the possibility of altering the set of rules  during the game. In particular, we propose a variant of the GoL with space of possible rules governed by rule modification processes. We consider two scenarios -- probabilistic rule switching and deterministic rule switching. Our goal is to demonstrate the trade-off between the stability and the growth in the system. We provide arguments suggesting that this trade-off can be controlled if one can consider synchronous and asynchronous updating policies combined with the rule selection mechanisms.

The remaining part of this paper is organized in the following manner. In Section~\ref{sec:model}, we extend the standard model of the GoL and  introduce two variants of the rule switching mechanisms by enabling the selection of the rules during the evolution. We investigate the properties of the introduced models in Section~\ref{sec:results} by presenting a series of numerical results and commenting on the presented results. Finally, in Section~\ref{sec:final} we summarize the paper, discuss the possible extensions, provide possible applications of the selected models, and include some concluding remarks.

\section{Rule switching mechanisms}\label{sec:model}

The standard version of the GoL is defined by the 2D lattice, with each cell being in one of the states -- dead (state 0, white cell) or alive (state 1, black cell). At each step, the cell is updated by taking into account its Moore neighbourhood. If the living cell has two or three living neighbours, it will stay alive. However, if it has only one living neighbour (underpopulation) or more than three living neighbours (overpopulation) it will die. On the other hand, if a dead  cell has exactly three living neighbours, it will become alive, emulating the birth process. Finally, the state of the lattice is updated after the state of each cell has been calculated.

Clearly, the requirement for updating the start of all cells at the same time is very limiting. Thus, it is natural to consider a relaxed version of the GoL, where only a fraction of cells has their state updated simultaneously~\cite{blok1999synchronous}. However, the rule used by each cell is independent and identical at each step, which limits the flexibility of the model.

This limitation is intrinsic to deterministic cellular automata. To address it, a stochastic version of 1D version of the GoL was proposed in \cite{monetti1997stochastic}. In this case, the randomness is incorporated into the model by altering the rule governing the evolution of each cell. In particular, in 1D model from \cite{monetti1997stochastic}, a cell has a fixed probability of remaining alive in the case of overpopulation.

In this work, we aim at combining both approaches, and we consider an extended version of the GoL with the possibility of choosing at each step a rule to be used for updating the state. Thus, we extend the freedom of the system by the possibility of altering the rules used at each step to calculate the next state of the cells. Additionally, we consider two modes of updating. In the asynchronous mode, each cell updates its state immediately after calculating it, while in the synchronous mode, the state update is the final element of the simulation step.

As the GoL was designed to mimic the behaviour of living organisms such as bacteria, it is natural to expect that the choice of rules will be dictated by the desire to balance the interplay between the resistance to overpopulation and the need to breed new living cells. In the presented work we focus only on the changes dictated by the alternation of the cell resistance to overpopulation.

For the sake of simplicity, we do not consider the memory of the process. To be more specific, in the described models the rules selected at each step do not depend on the history of previous decisions. In such case, it is natural to consider two possible scenarios.

In the memory-less scenario, the rule to be applied at step is based on a random choice made during the move. To be more precise, the random choice is used to select the rule-set to be used. As we are aiming at using the modified definition of the GoL, the particular rule for each cell depends -- as in the case of the standard GoL -- on the number of living cells in its neighbourhood.

Alternatively, one can consider a scenario in which the rule switching is based on a prescribed sequence of moves. In this case, the resulting game is equivalent to a deterministic composition of two games. Again, in this variant, one can consider a variety of processes used to construct the policy of the cells for choosing the update rules. In this paper, we consider only a fully deterministic case, where the rules from some set of rules are used periodically.

\subsection{Probabilistic rule switching}
Let us start by considering the case where the rule selection is  governed by a memoryless, random process. In such mode we extend the original GoL and include two evolution rules, $\golRule{1}$ and $\golRule{2}$  Each of the rules is  given with a fixed probability, $p$ and $q=1-p$, respectively. Additionally, we assume that the probability of choosing one of the rules is constant during the evolution.

As it has already been pointed out, we restrict our considerations to the case where the behaviour of the system is altered with respect to the overpopulation. This is equivalent to stating that we consider the case where the rules differ regarding the threshold, $r$, for which the living cell will die due to the overcrowding. In the case of the standard GoL, this threshold is set to $r=4$. Such setup leads to the process described by the pseudocode presented in \procName~\ref{proc:gol-radnom-rules}.

\begin{algorithm}[H]
\begin{algorithmic}
    \REQUIRE threshold $r$ for overpopulation, probability $p$ of using the standard rule
    \FOR{$i=0$ to $n$} 
        \STATE $x$ $\leftarrow$ U(0,1) \COMMENT{Random number from uniform distribution on [0,1)}
    \IF{$x \geq p$} 
      \STATE $c_i \leftarrow$ $GoL(2,3,r)$ \COMMENT{Use an alternative rule}
    \ELSE
      \STATE $c_i \leftarrow$ $GoL(2,3,4)$ \COMMENT{Use the standard rule, $r=4$}
    \ENDIF
    \ENDFOR
\end{algorithmic}
\caption{Random selection of rules in the GoL}
\label{proc:gol-radnom-rules}
\end{algorithm}

Clearly, the situation presented in Procedure~\ref{proc:gol-radnom-rules} describes one of the possible situations where each cell can use one of the rules from a predefined set of rules. One can consider more complex methods, \eg with more alternative rules used in the procedures.

The process of selecting the rule to be used at each step is governed by the uniform distribution. However, in a general case one could incorporate into this model an arbitrary strategy for randomly selecting one of the rules, $\golRule{1}, \golRule{2},\dots, \golRule{n}$, from a predefined set. This, in principle, can be used to alter the behaviour of the cells by modifying their preferences for choosing the rules.

\subsection{Deterministic rule switching}

In the above scenario, the limitation of the standard GoL for using only a predefined rule was lifted by introducing random selection policy. However, the rule switching mechanism can be introduced in many different manners.

As an alternative approach, one can consider a deterministic policy for switching between the rules. In such case, instead of selecting among the available rules, $\golRule{1}, \golRule{2},\dots, \golRule{n}$, the system is evolving by specifying the order of utilizing the rules at each step, $\golRule{k_1}^{1}\golRule{k_2}^{1}\dots\golRule{k_m}^{m}$. 

A particulate case of this scheme, with $n=2$, is presented in  Procedure~\ref{proc:gol-deterministic-alternting}.
Here, we restrict ourselves to  $\golRule{1}$ and $\golRule{2}$, representing the standard rule and the rule with the modified resistance for the overpopulation. In this particular case, the specification of the rule switching is given by the string of 0s and 1s.
 
\begin{algorithm}[H]
\begin{algorithmic}
    \REQUIRE threshold $r$ for the alternative rule, $s$ string of $0$s and $1$s describing the rule selection mechanism
    \FOR{$i=0$ to $n$} 
        \STATE $x \leftarrow s[i \pmod{ \|s\|}]$ \COMMENT{Select the next element of $s$}
    \IF{$x = 1$} 
      \STATE $c_i \leftarrow$ $GoL(2,3,r)$
    \ELSE
      \STATE $c_i \leftarrow$ $GoL(2,3,4)$
    \ENDIF
    \ENDFOR
\end{algorithmic}
\caption{Deterministic selection of rules in the GoL based on the predefined sequence.}
\label{proc:gol-deterministic-alternting}
\end{algorithm}

One should note that the particular case of $n$ ones results in the situation where the alternative rule is always applied. Similarly, the sequence of $n$ zeros leads to the standard version of the GoL. On the other hand, it is possible to define an arbitrary policy for rule selection by specifying a string $s$ or by defining a method for updating it depending on the internal state of the system or some external conditions. Thanks to this, the proposed version of the GoL is more flexible in the applications related to biological and social systems, where it is natural to expect that the rule selection policy will evolve due to some factors affecting the system.

\subsection{Updating policy}
In addition to considering the possibility of switching rules during the course of the game, we consider two cases of the updating policy. In the synchronous case, the state of each cell is calculated and the state updating is executed after all cells have their next state calculated. Naturally, in this case, the order of calculating the next state is not relevant. 

On the other hand, we also consider a situation where the state of the cell is updated immediately after it is calculated. Thus, the order of the cells used to calculate their next state is important~\cite{comer2014who}. In particular, one should note that the updating order could also be used to significantly alter the behaviour of the model. For example, in the population with two variants of agent, the updating speed could be different. However, in the following considerations, we assume that the cells are selected in a random order.

\section{Numerical results}\label{sec:results}

In this section we present the results of numerical experiments executed for  selected instances of the modified GoL. We start with the presentation of the patters forming when the variants introduced in the previous section are used. For this purpose, we include a comparison of the final configurations obtained for different setups. Next, we describe the stability of the system in synchronous and asynchronous scenarios. Finally, we show to what  degree the change in the rule selection mechanism and the synchronization policy has impact of the growth in the system.

\subsection{Pattern formation}\label{sec:results-patterns}

\begin{figure*}[t!]
	\centering
	Synchronous updating\\
	\subfigure[$r=4,p=0.25$]{\includegraphics[width=0.18\textwidth]{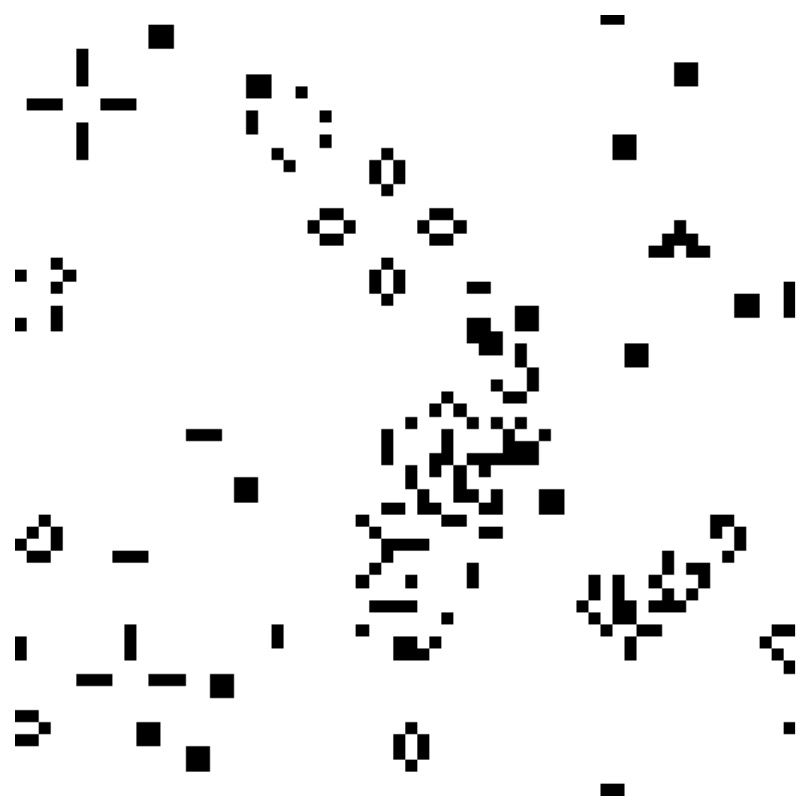}\label{fig:patern-formation-s-4-025}}
	\subfigure[$r=5,p=0.25$]{\includegraphics[width=0.18\textwidth]{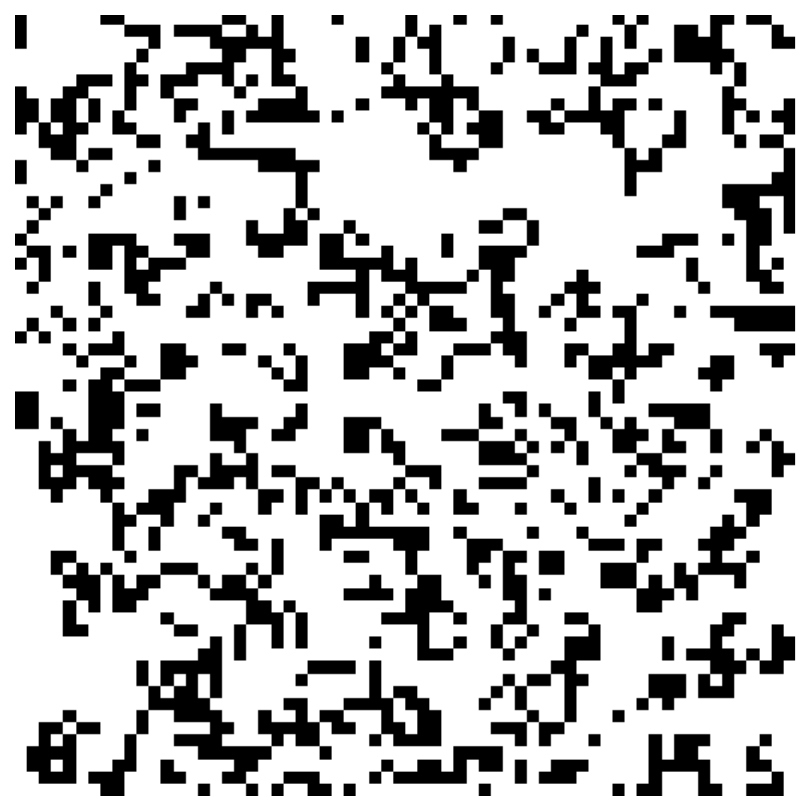}\label{fig:patern-formation-s-5-025}}
	\subfigure[$r=6,p=0.25$]{\includegraphics[width=0.18\textwidth]{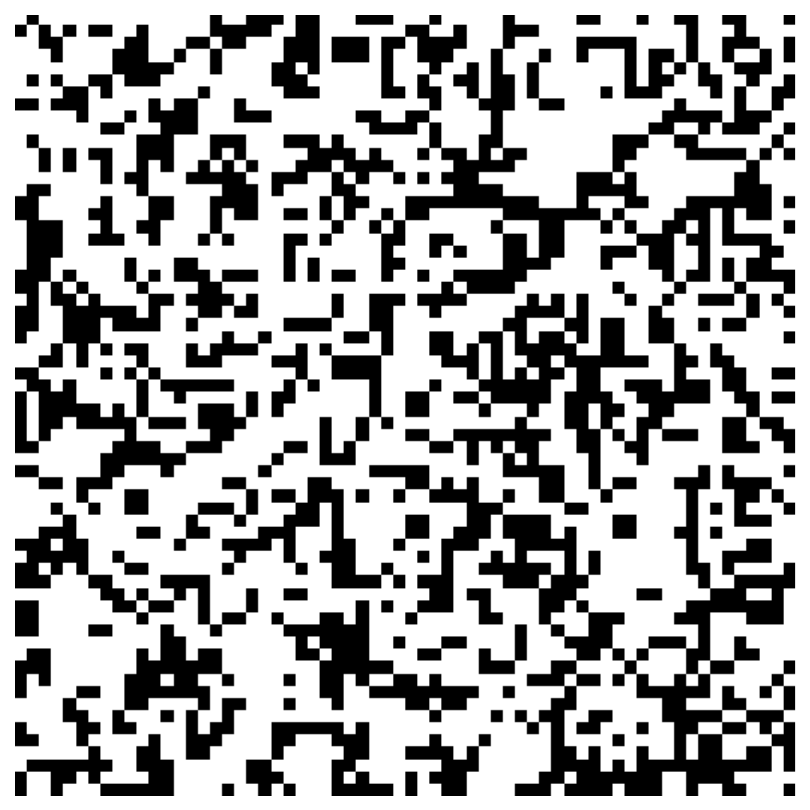}\label{fig:patern-formation-s-6-025}}
	\subfigure[$r=7,p=0.25$]{\includegraphics[width=0.18\textwidth]{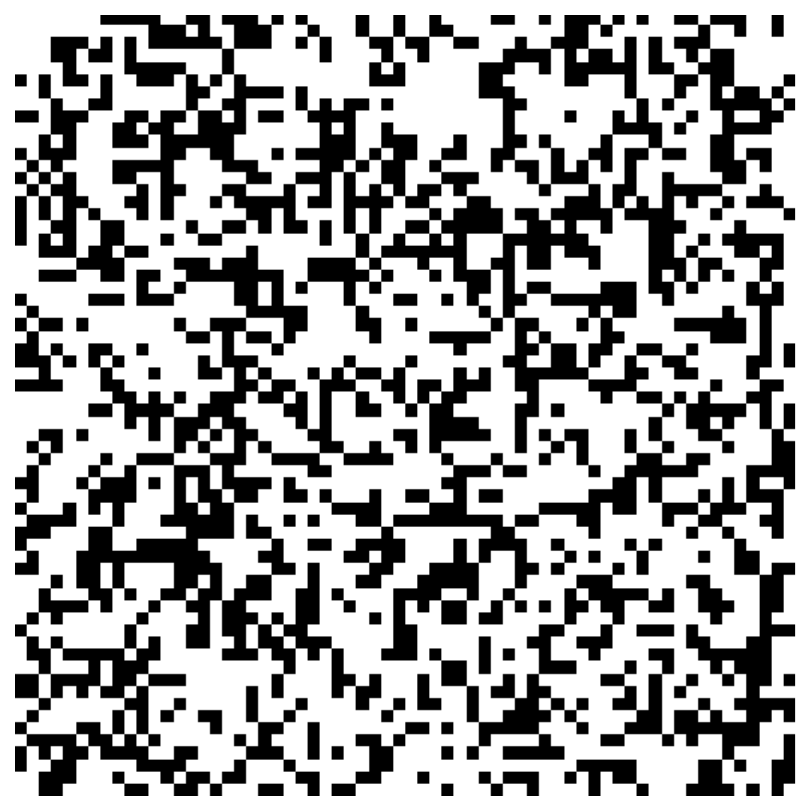}\label{fig:patern-formation-s-7-025}}
	\subfigure[$r=8,p=0.25$]{\includegraphics[width=0.18\textwidth]{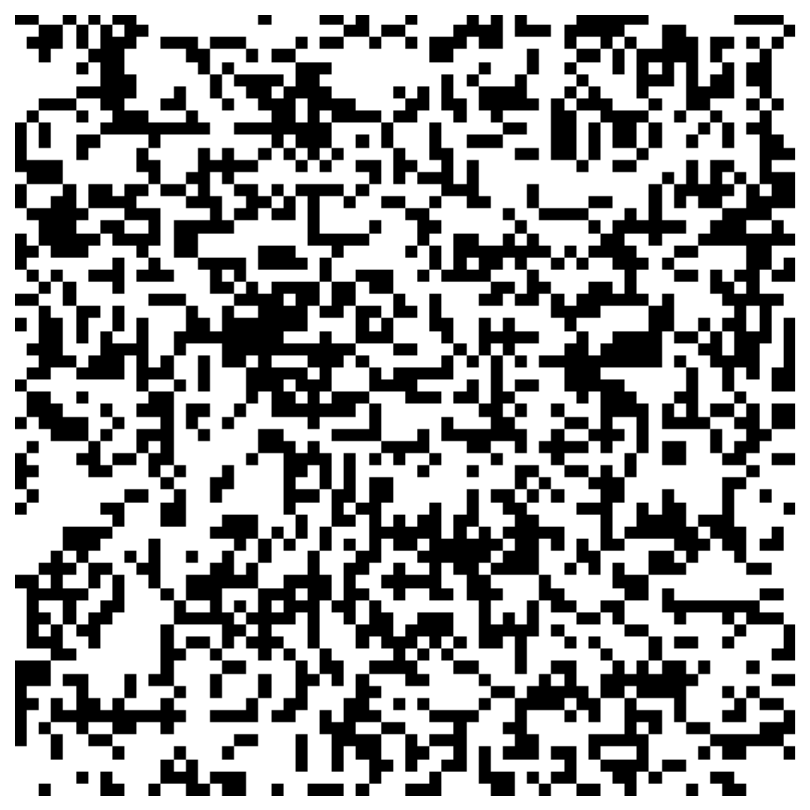}\label{fig:patern-formation-s-8-025}}
	
	\subfigure[$r=4,p=1$]{\includegraphics[width=0.18\textwidth]{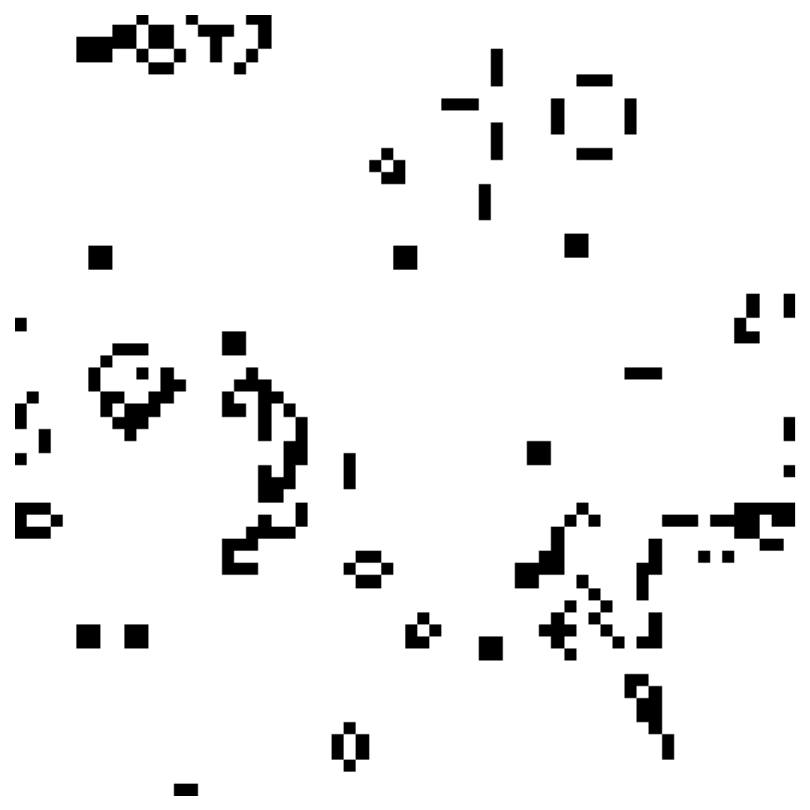}\label{fig:patern-formation-s-4-1}}
	\subfigure[$r=5,p=1$]{\includegraphics[width=0.18\textwidth]{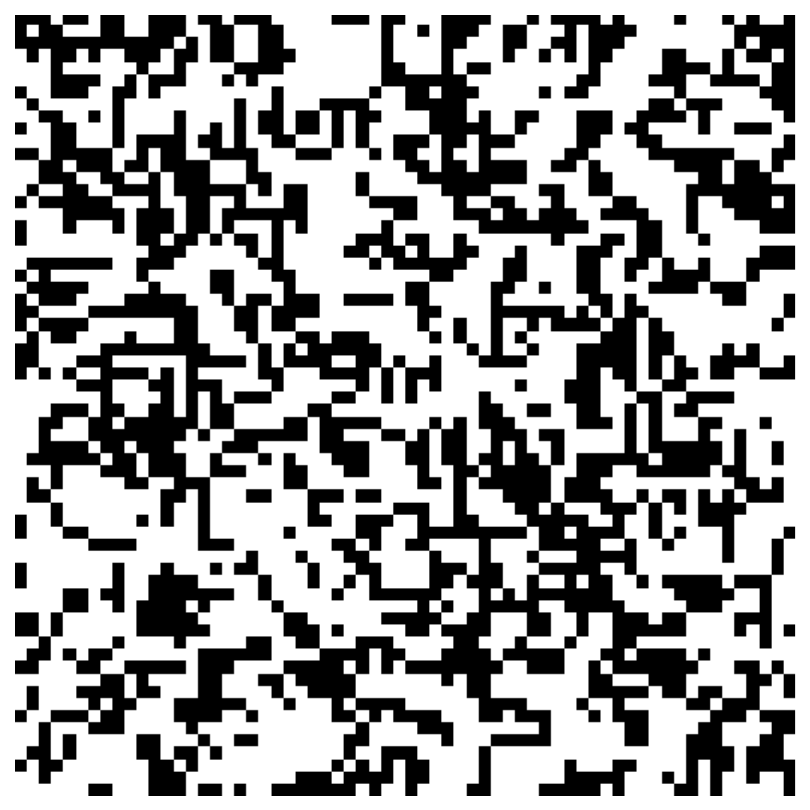}\label{fig:patern-formation-s-5-1}}
	\subfigure[$r=6,p=1$]{\includegraphics[width=0.18\textwidth]{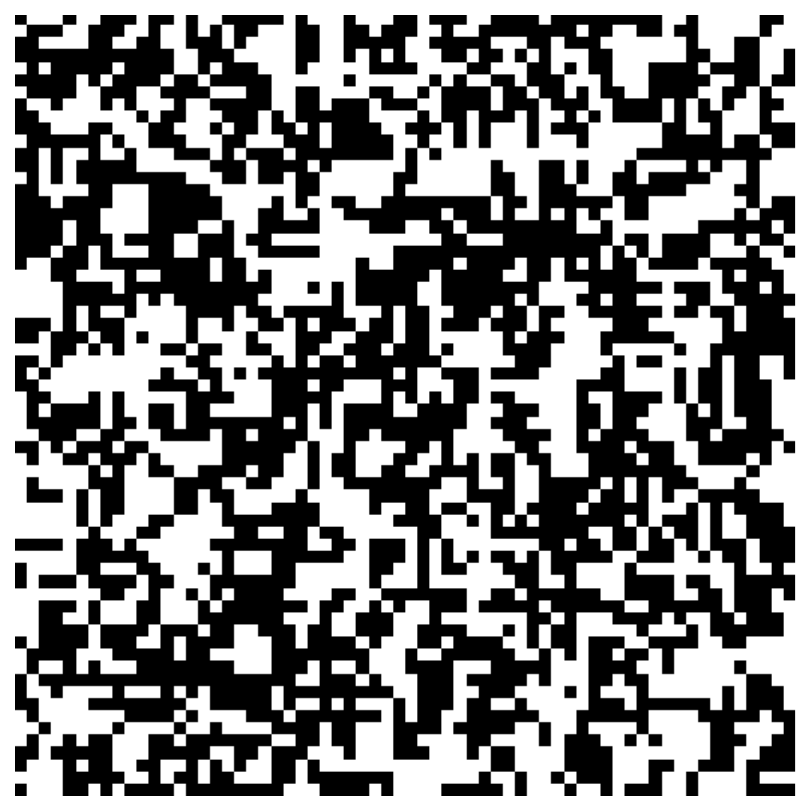}\label{fig:patern-formation-s-6-1}}
	\subfigure[$r=7,p=1$]{\includegraphics[width=0.18\textwidth]{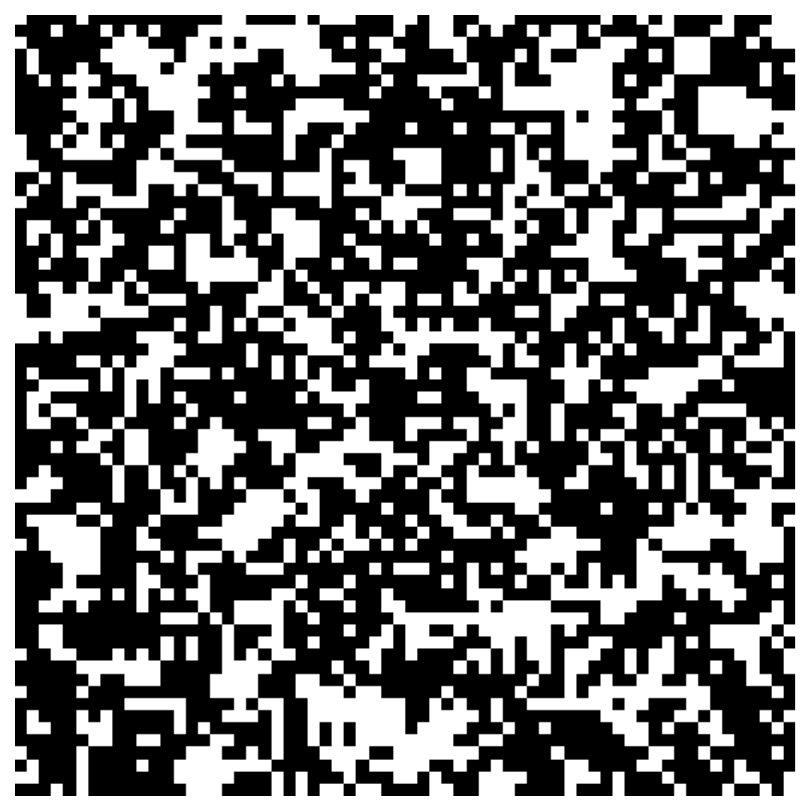}\label{fig:patern-formation-s-7-1}}
	\subfigure[$r=8,p=1$]{\includegraphics[width=0.18\textwidth]{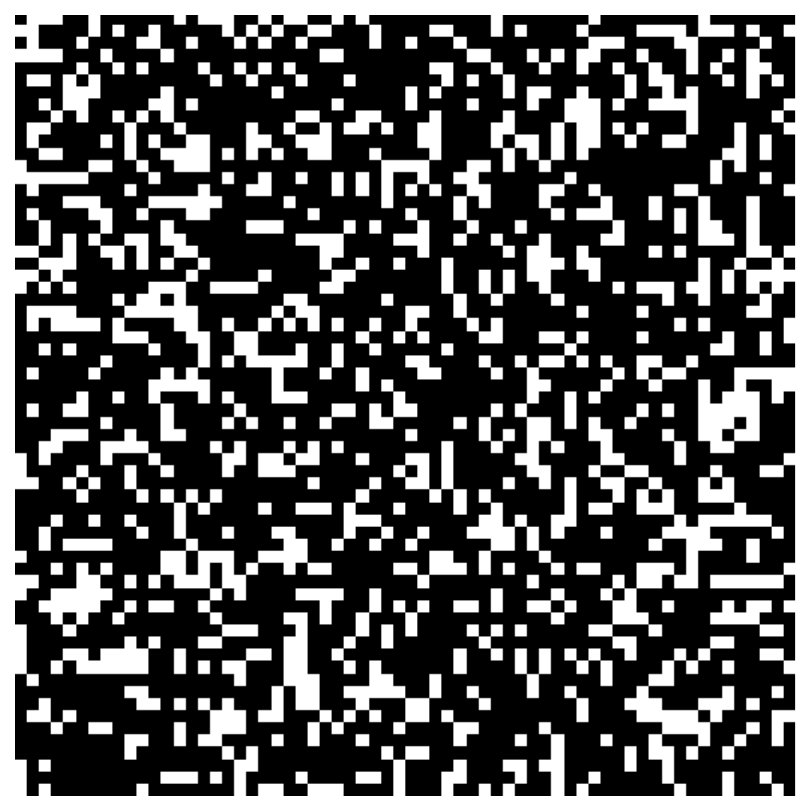}\label{fig:patern-formation-s-8-1}}
	
	Asynchronous updating\\
	\subfigure[$r=4,p=0.25$]{\includegraphics[width=0.18\textwidth]{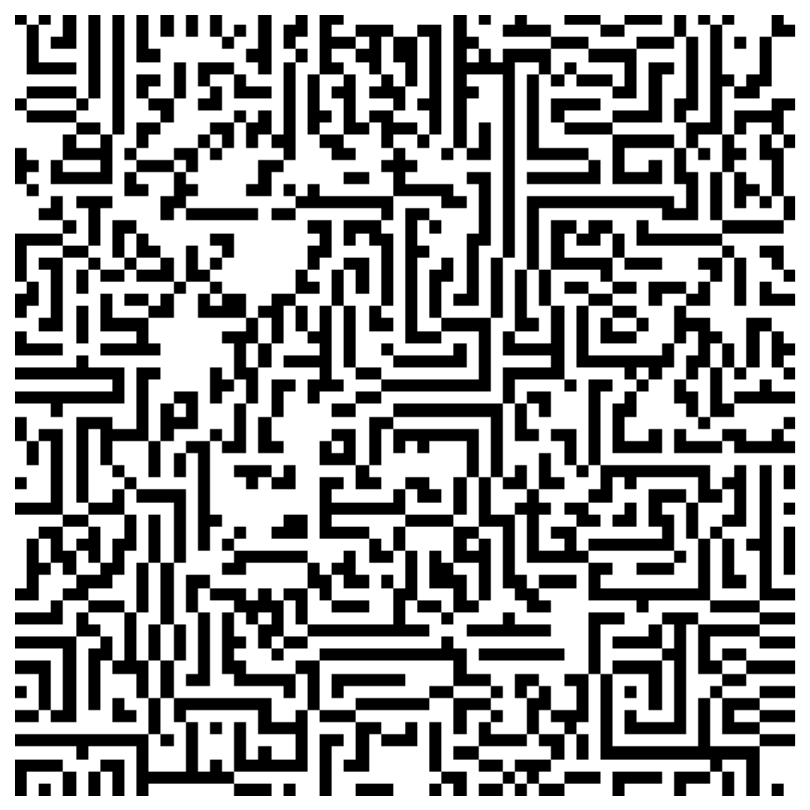}\label{fig:patern-formation-a-4-025}}
	\subfigure[$r=5,p=0.25$]{\includegraphics[width=0.18\textwidth]{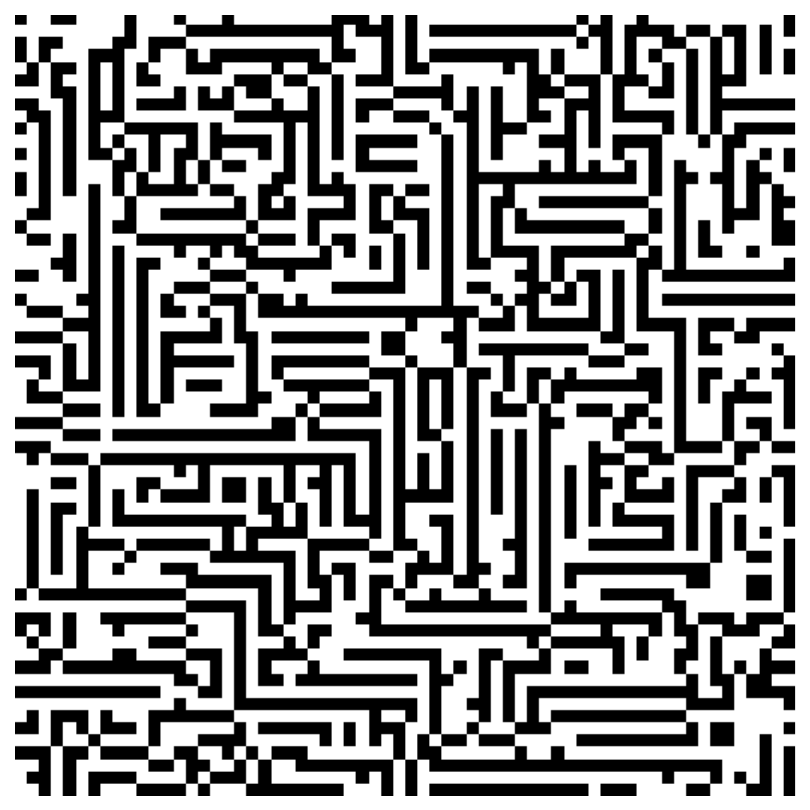}\label{fig:patern-formation-a-5-025}}
	\subfigure[$r=6,p=0.25$]{\includegraphics[width=0.18\textwidth]{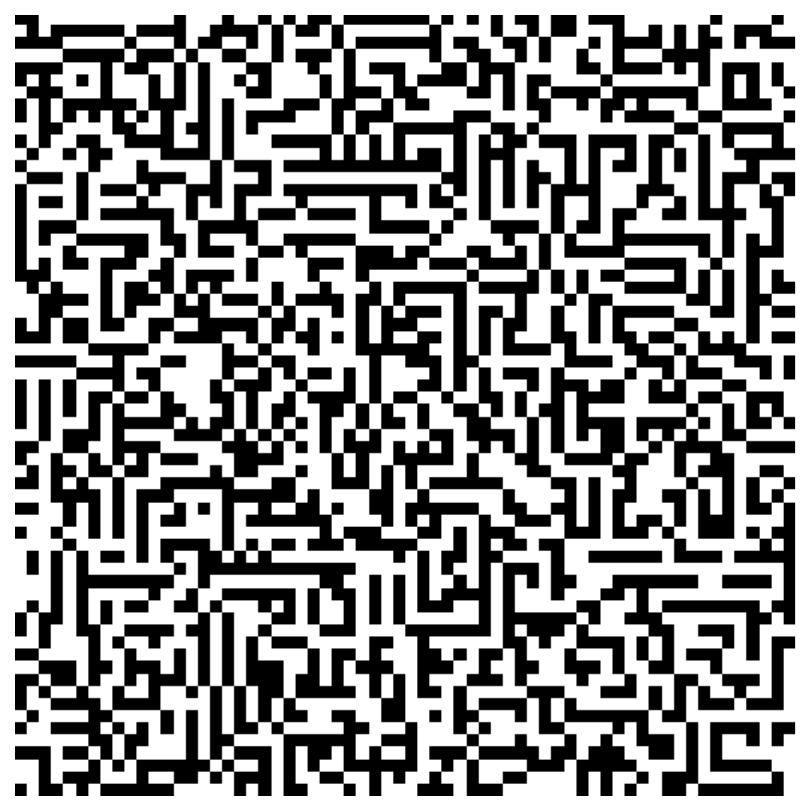}\label{fig:patern-formation-a-6-025}}
	\subfigure[$r=7,p=0.25$]{\includegraphics[width=0.18\textwidth]{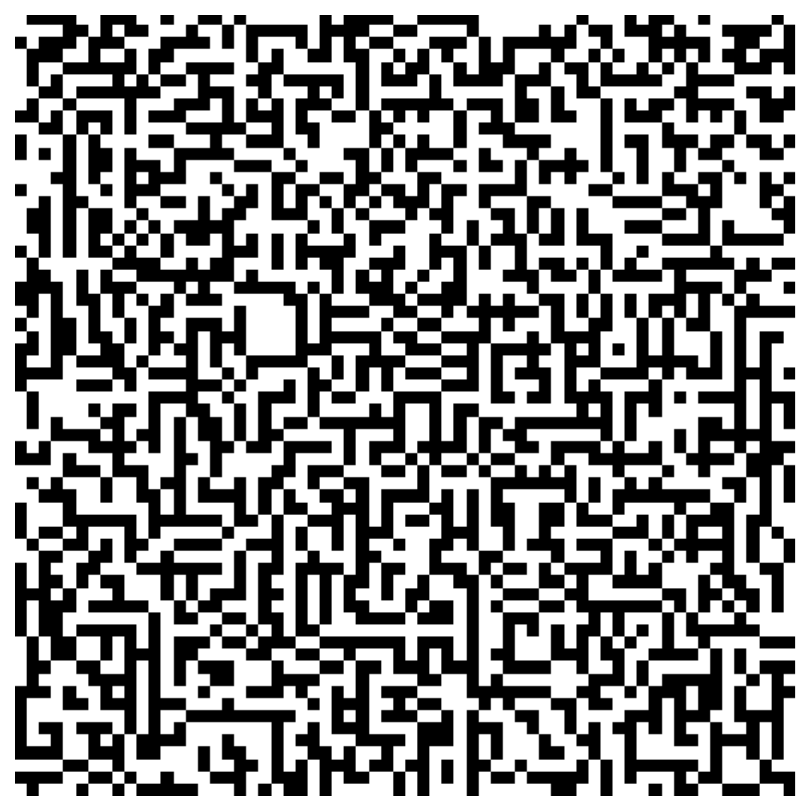}\label{fig:patern-formation-a-7-025}}
	\subfigure[$r=8,p=0.25$]{\includegraphics[width=0.18\textwidth]{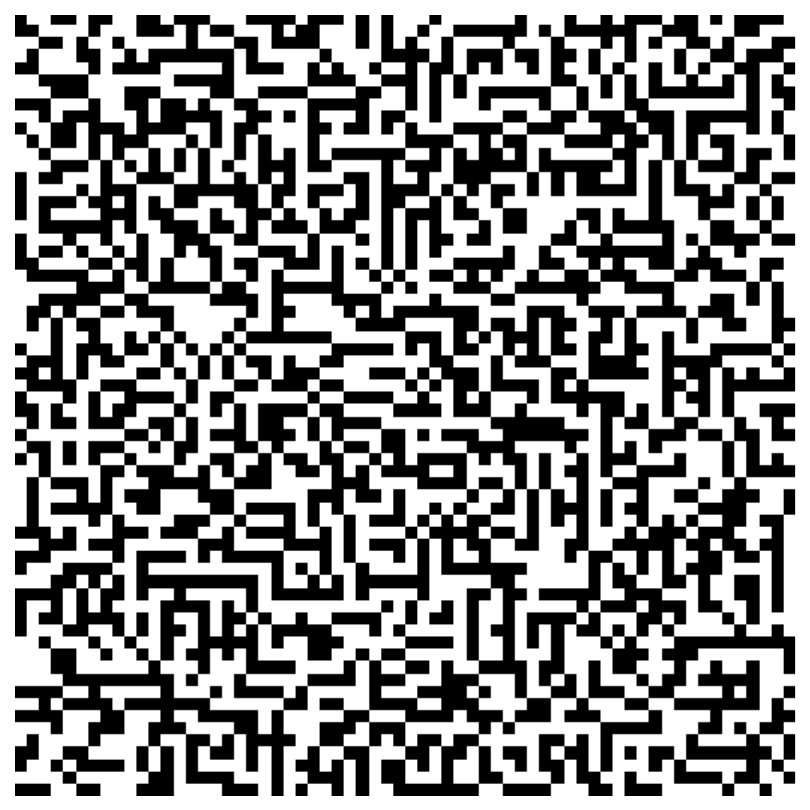}\label{fig:patern-formation-a-8-025}}

	\subfigure[$r=4,p=1$]{\includegraphics[width=0.18\textwidth]{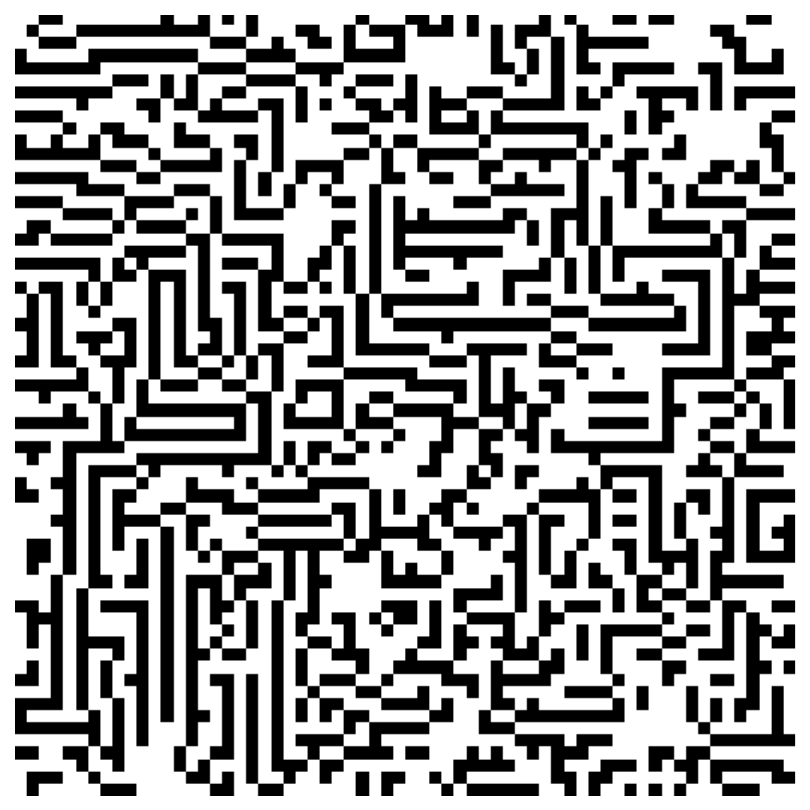}\label{fig:patern-formation-a-4-1}}
	\subfigure[$r=5,p=1$]{\includegraphics[width=0.18\textwidth]{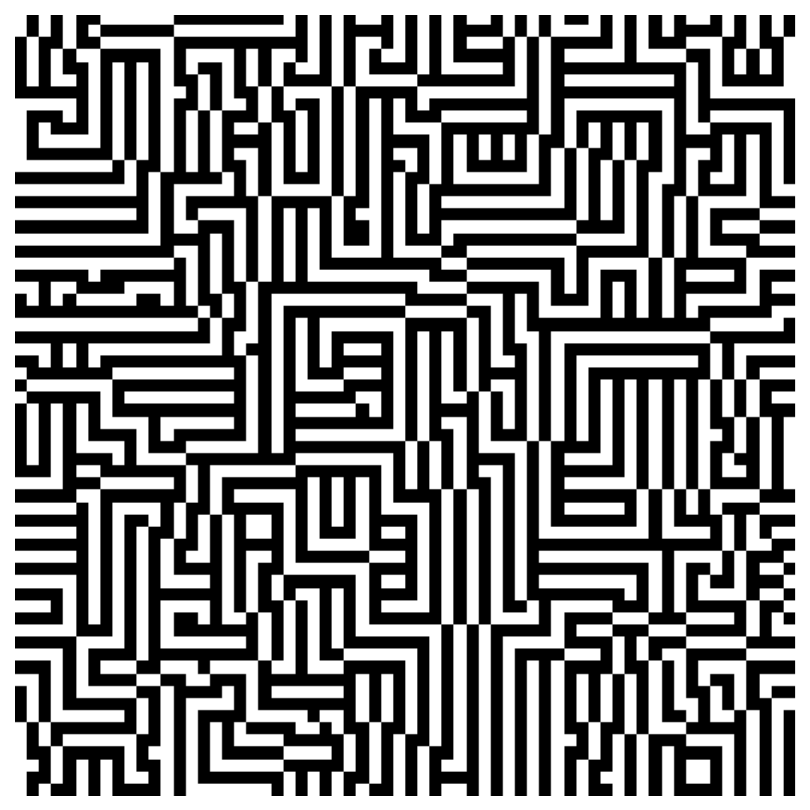}\label{fig:patern-formation-a-5-1}}
	\subfigure[$r=6,p=1$]{\includegraphics[width=0.18\textwidth]{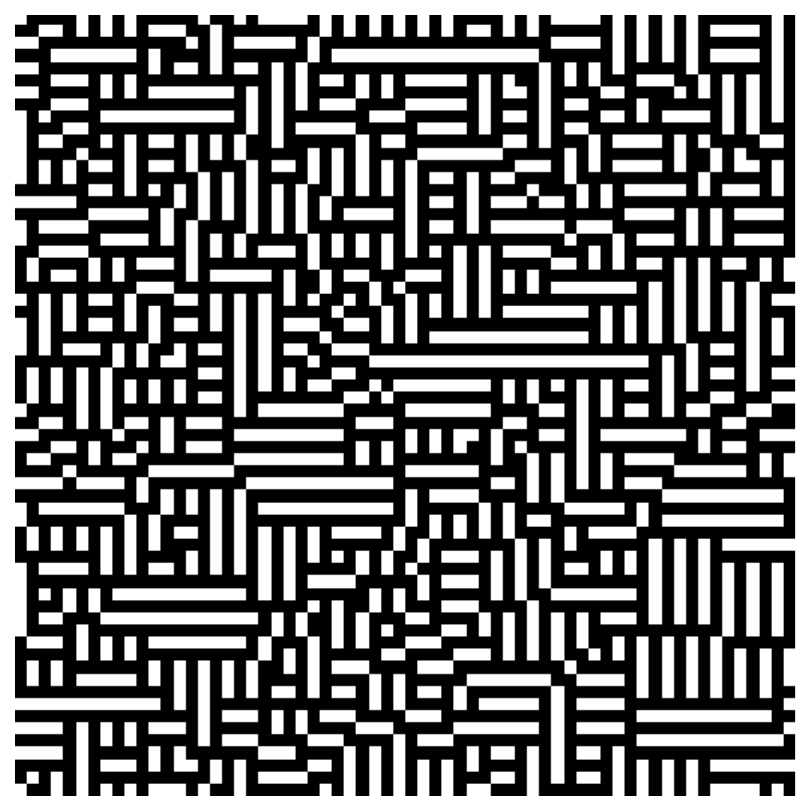}\label{fig:patern-formation-a-6-1}}
	\subfigure[$r=7,p=1$]{\includegraphics[width=0.18\textwidth]{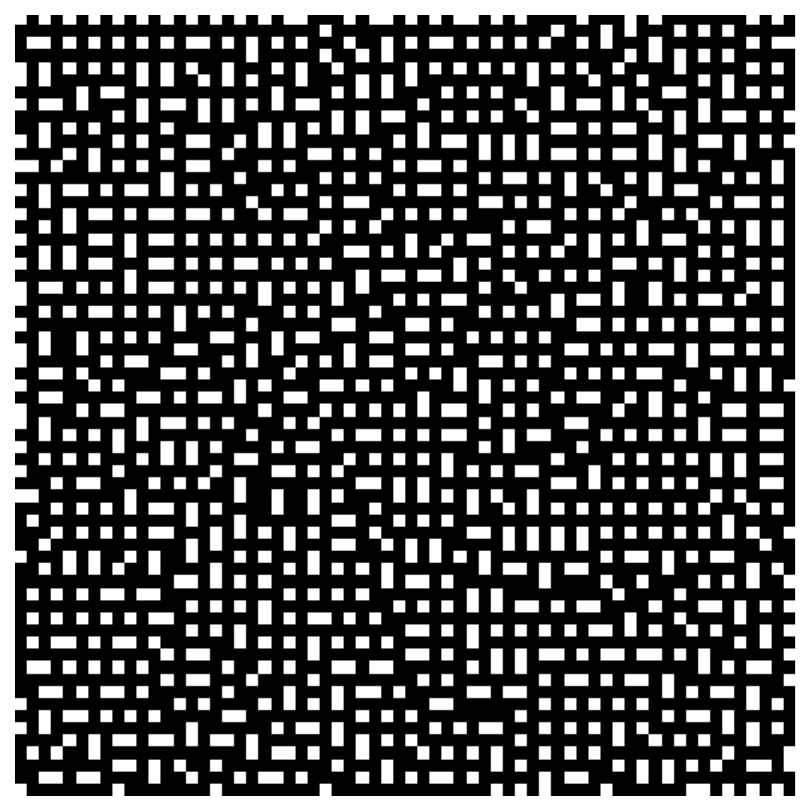}\label{fig:patern-formation-a-7-1}}
	\subfigure[$r=8,p=1$]{\includegraphics[width=0.18\textwidth]{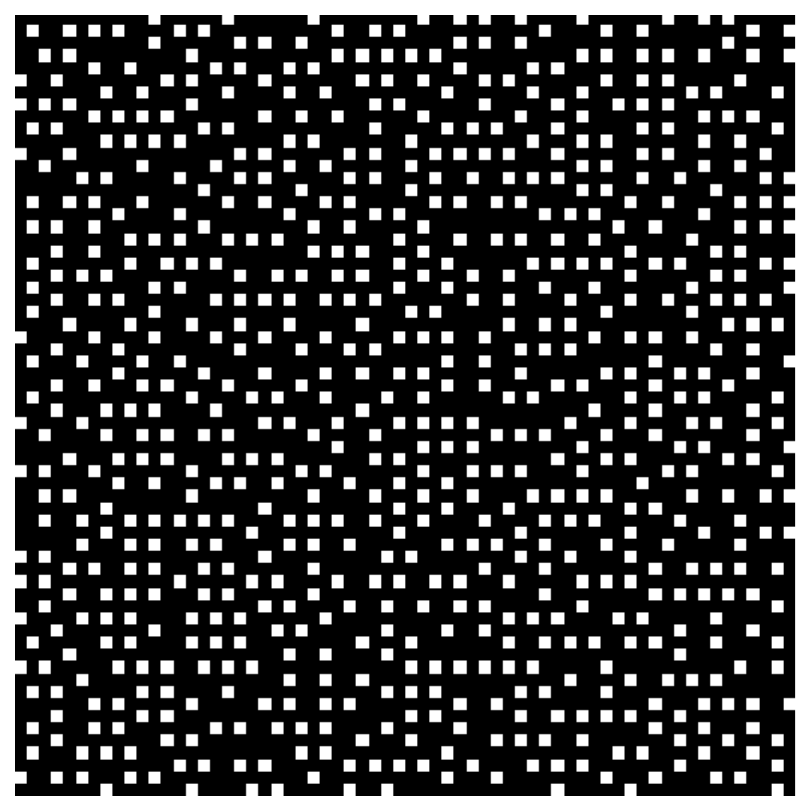}\label{fig:patern-formation-a-8-1}}
	
	\caption{Typical configurations for random rule switching mechanisms for the synchronous (top two rows) and the asynchronous (bottom two rows) updating policies. Examples of configurations after 500 steps of simulation for values of rule switching probability, $p=0.25$ and $p=1$, and the second threshold $r=4,5,6,7,8$ used in the alternative rule. The game was played on a periodic lattice of size $64\times 64$. In this case, the asynchronous updating leads to the formation of long-range patterns. The patters formed for the values of the threshold $r>4$ have homogeneity increasing with the value of $r$ and the probability of utilizing the rule based on the higher threshold. This effect does not occur for the synchronous updating, and one can conclude that the asynchronous policy leads to more predictable pattern formation.}
	\label{fig:patern-formation}
\end{figure*}

\begin{figure*}[t!]
\centering

Deterministic rule mechanism (alternating), synchronous updating\\
\subfigure[$r=4$]{\includegraphics[width=0.18\textwidth]{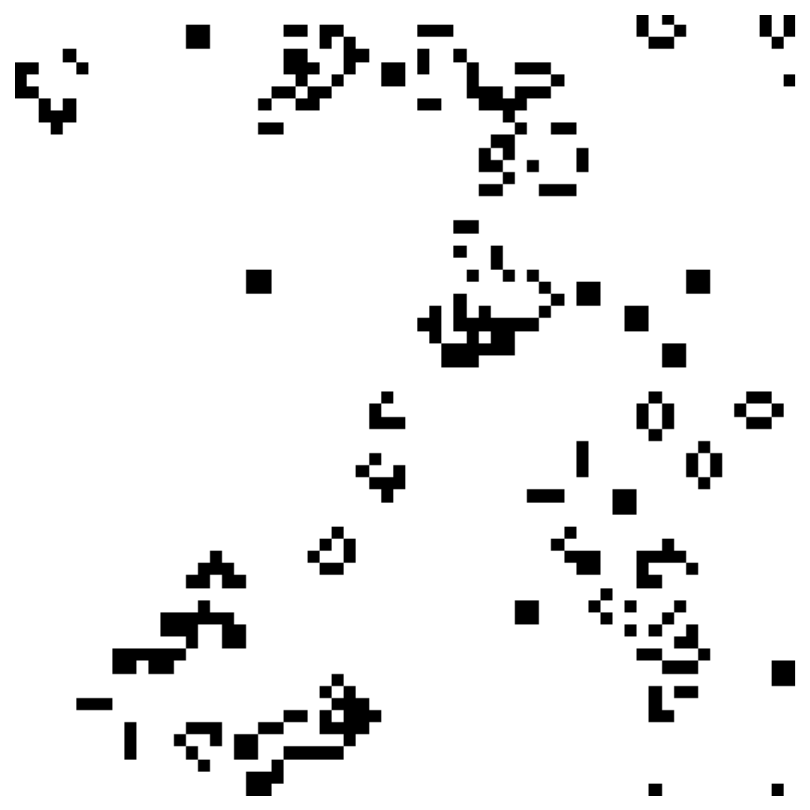}\label{fig:sa-rd-s-dp2-4}}
\subfigure[$r=5$]{\includegraphics[width=0.18\textwidth]{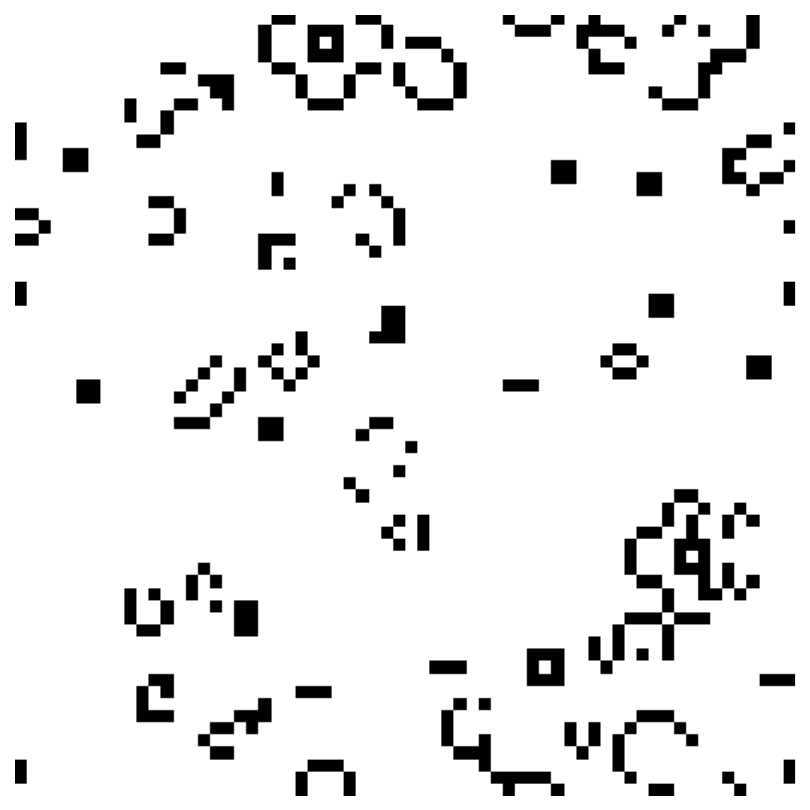}\label{fig:sa-rd-s-dp2-5}}
\subfigure[$r=6$]{\includegraphics[width=0.18\textwidth]{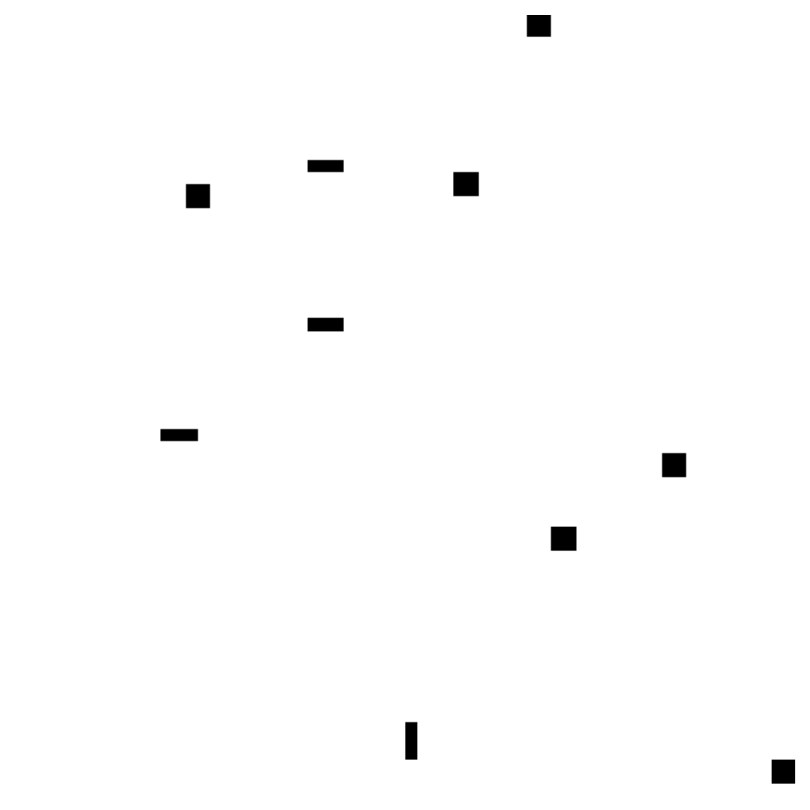}\label{fig:sa-rd-s-dp2-6}}
\subfigure[$r=7$]{\includegraphics[width=0.18\textwidth]{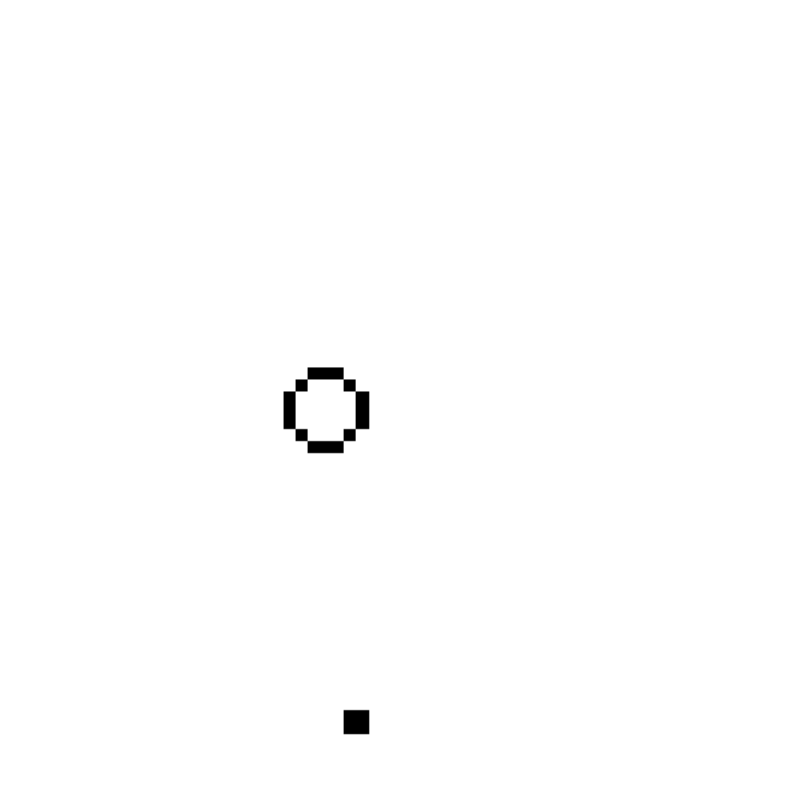}\label{fig:sa-rd-s-dp2-7}}
\subfigure[$r=8$]{\includegraphics[width=0.18\textwidth]{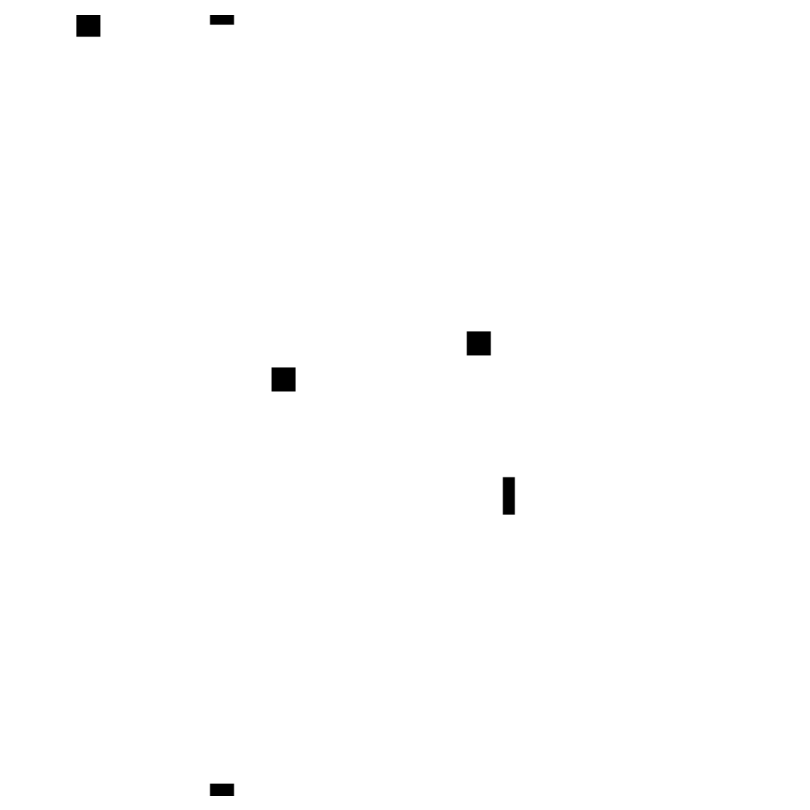}\label{fig:sa-rd-s-dp2-8}}

Deterministic rule mechanism (alternating), asynchronous updating\\
\subfigure[$r=4$]{\includegraphics[width=0.18\textwidth]{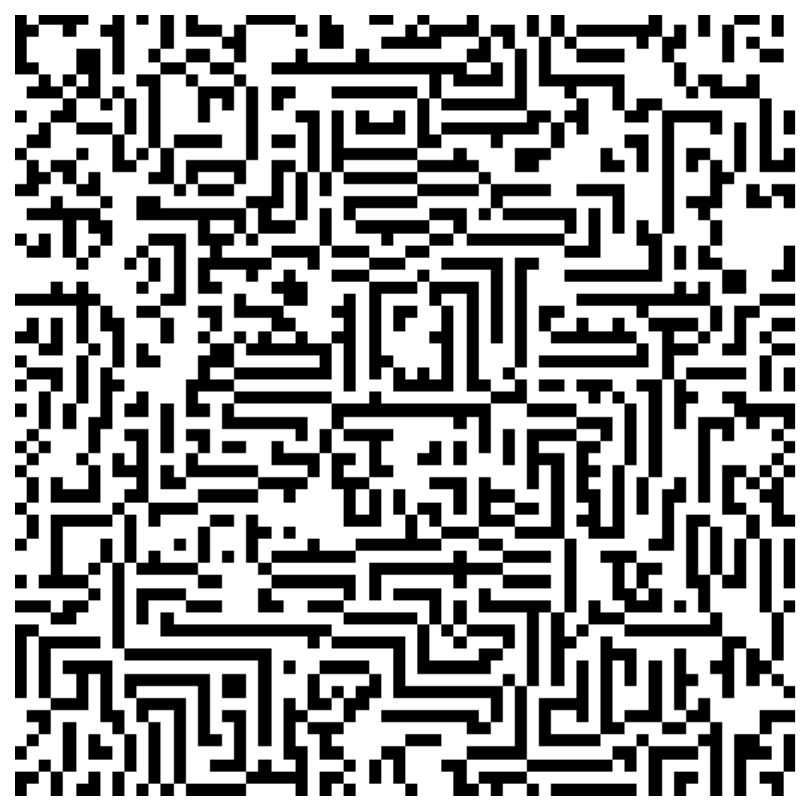}\label{fig:sa-rd-a-dp2-4}}
\subfigure[$r=5$]{\includegraphics[width=0.18\textwidth]{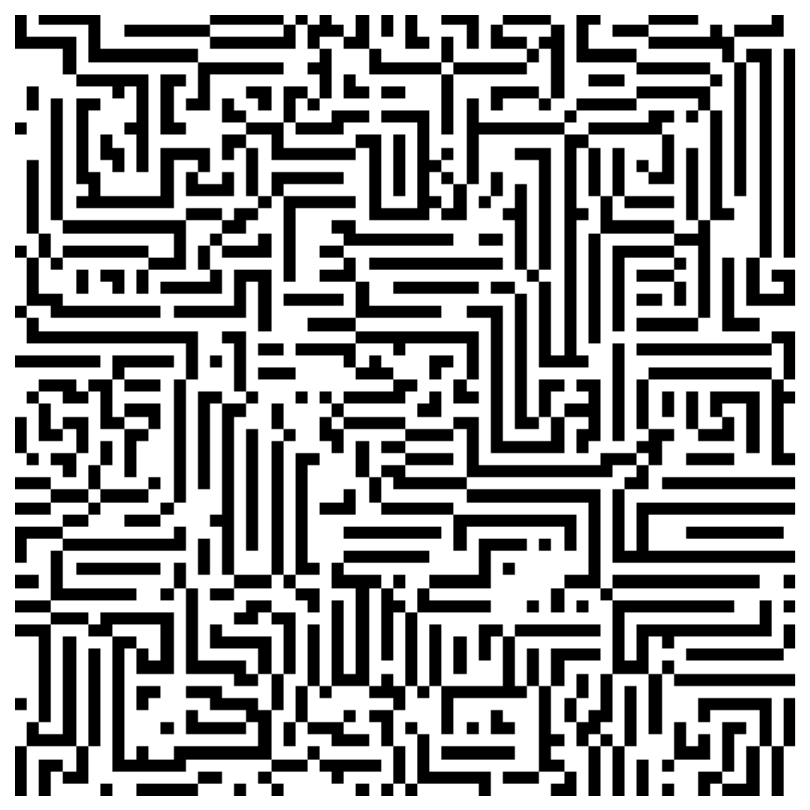}\label{fig:sa-rd-a-dp2-5}}
\subfigure[$r=6$]{\includegraphics[width=0.18\textwidth]{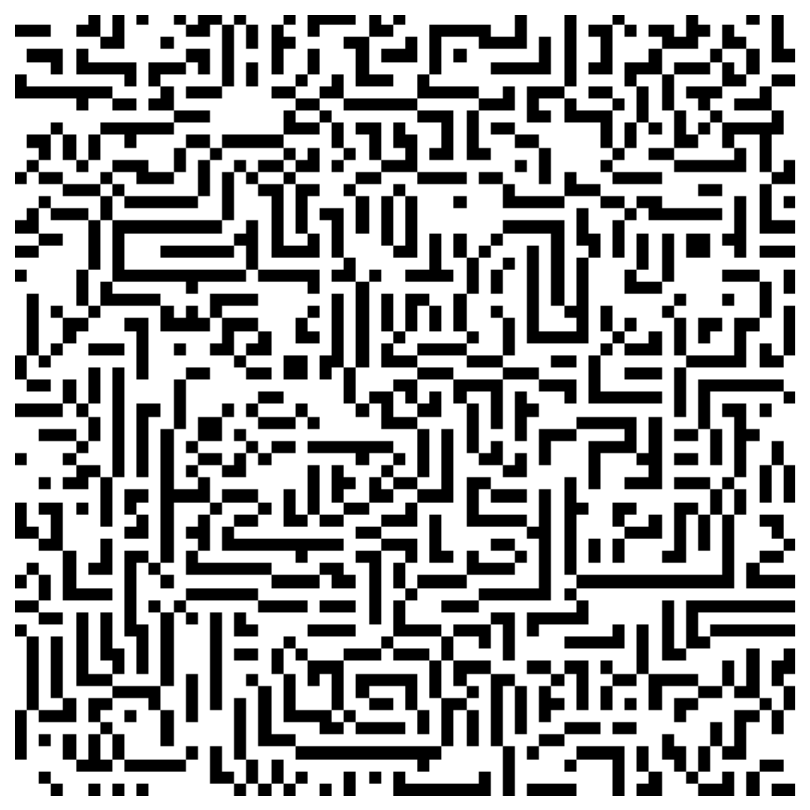}\label{fig:sa-rd-a-dp2-6}}
\subfigure[$r=7$]{\includegraphics[width=0.18\textwidth]{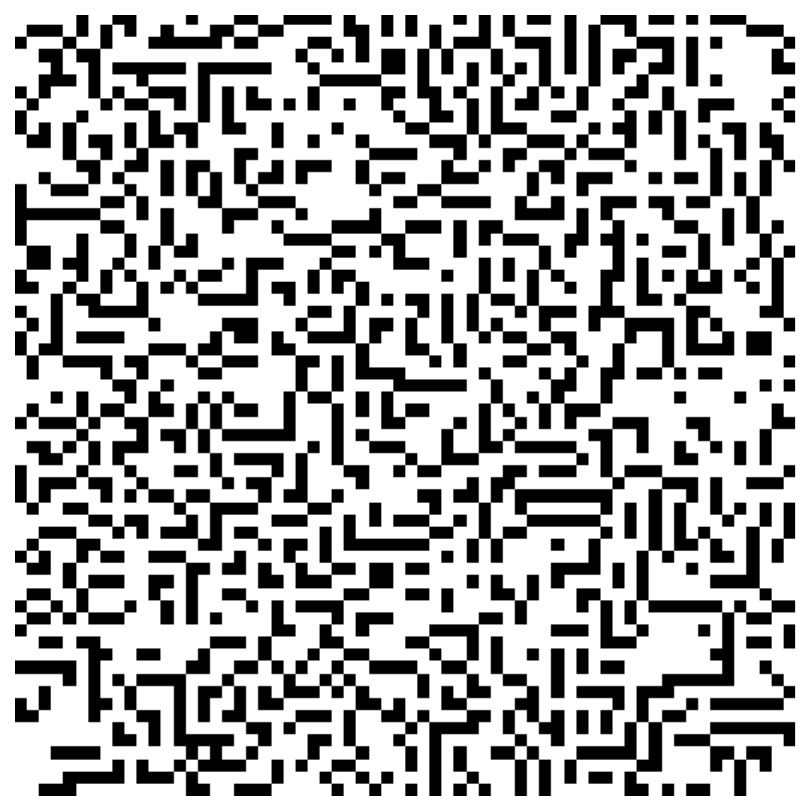}\label{fig:sa-rd-a-dp2-7}}
\subfigure[$r=8$]{\includegraphics[width=0.18\textwidth]{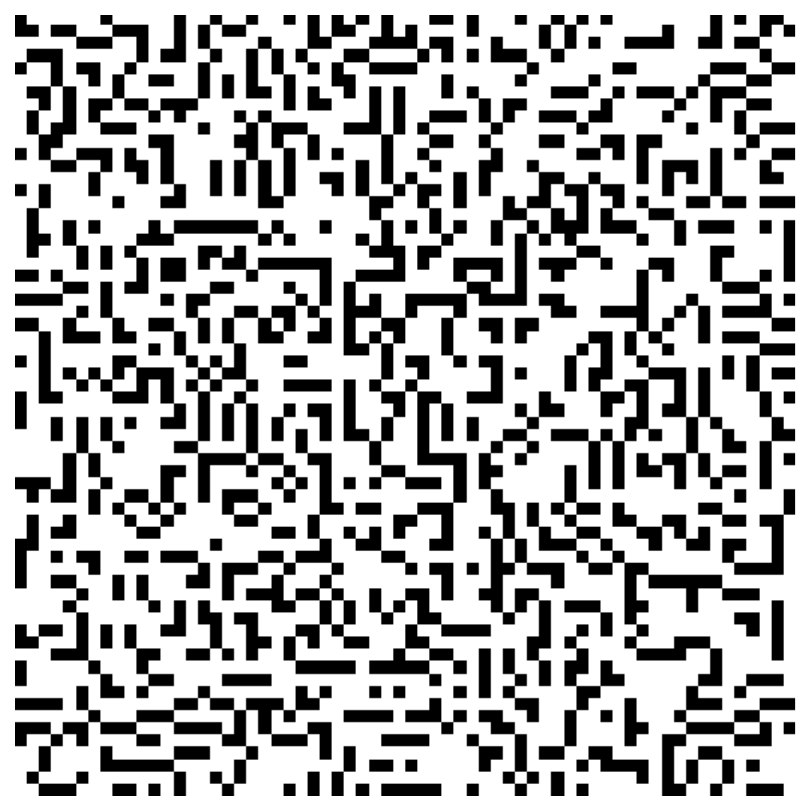}\label{fig:sa-rd-a-dp2-8}}

Random rule mechanism ($p=0.5$), synchronous updating \\
\subfigure[$r=4$]{\includegraphics[width=0.18\textwidth]{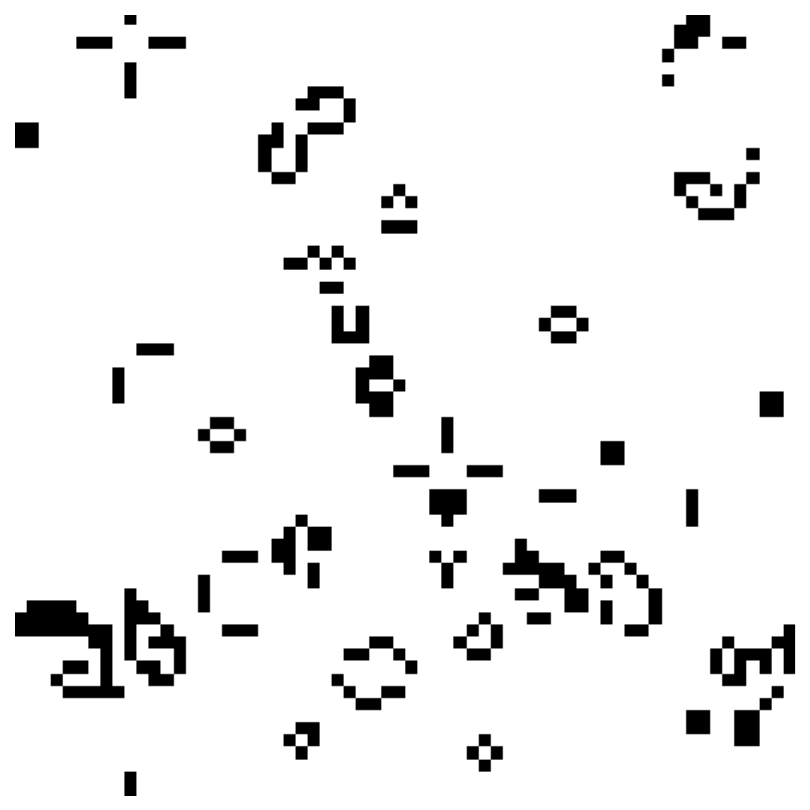}\label{fig:sa-rd-s-r-4}}
\subfigure[$r=5$]{\includegraphics[width=0.18\textwidth]{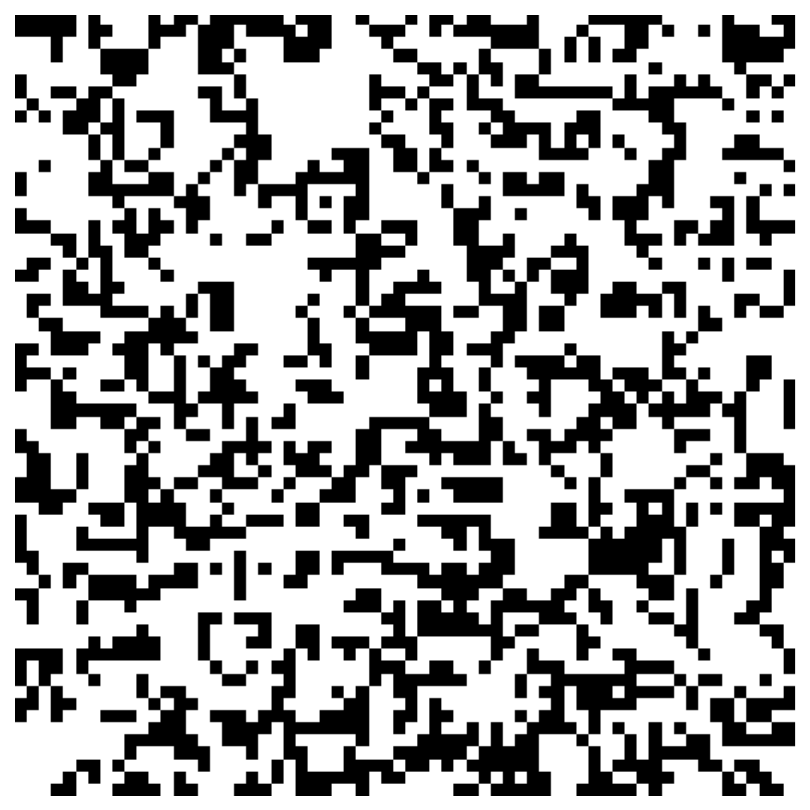}\label{fig:sa-rd-s-r-5}}
\subfigure[$r=6$]{\includegraphics[width=0.18\textwidth]{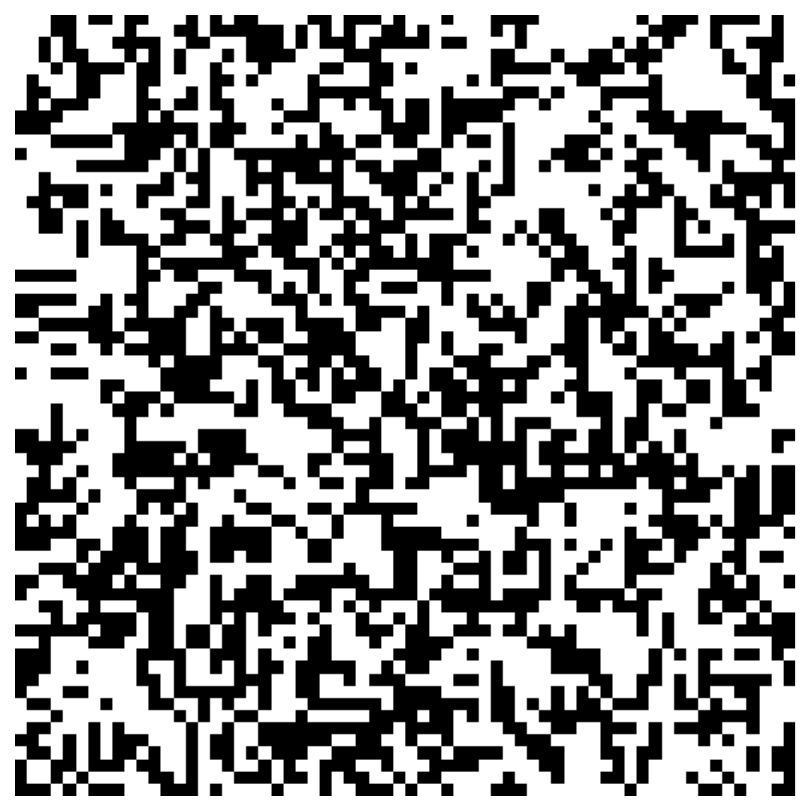}\label{fig:sa-rd-s-r-6}}
\subfigure[$r=7$]{\includegraphics[width=0.18\textwidth]{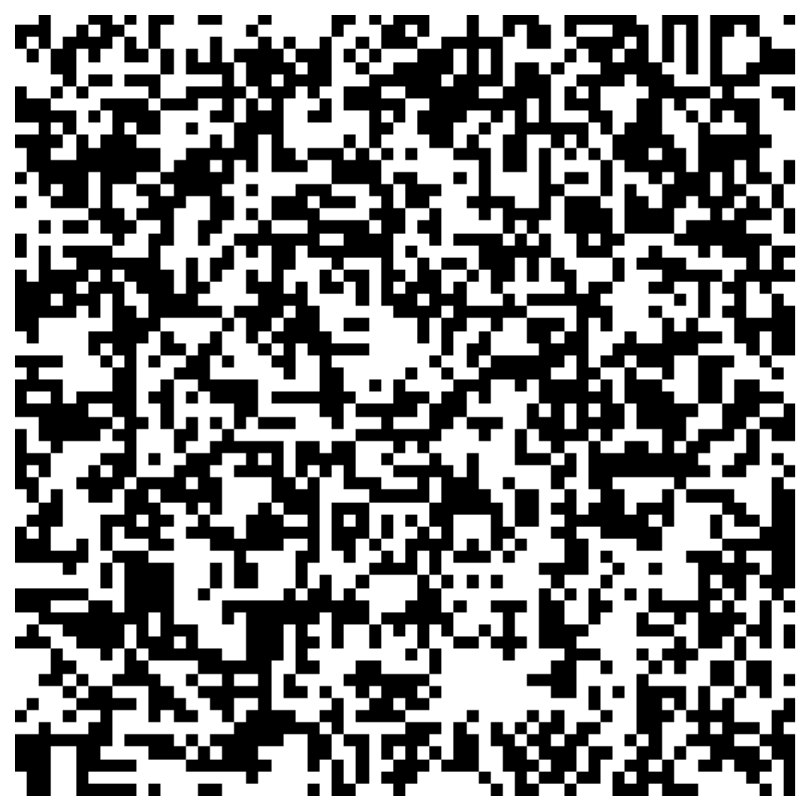}\label{fig:sa-rd-s-r-7}}
\subfigure[$r=8$]{\includegraphics[width=0.18\textwidth]{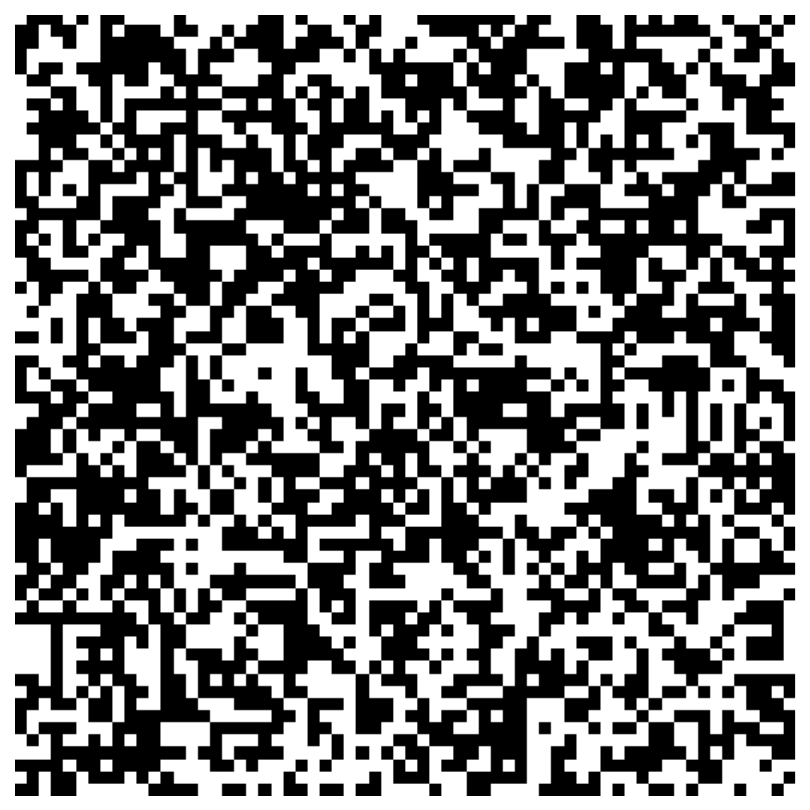}\label{fig:sa-rd-s-r-8}}

Random rule mechanism ($p=0.5$), asynchronous updating\\
\subfigure[$r=4$]{\includegraphics[width=0.18\textwidth]{plot_world_size64_init50_sync_rand_p0.50_st4.pdf}\label{fig:sa-rd-a-r-4}}
\subfigure[$r=5$]{\includegraphics[width=0.18\textwidth]{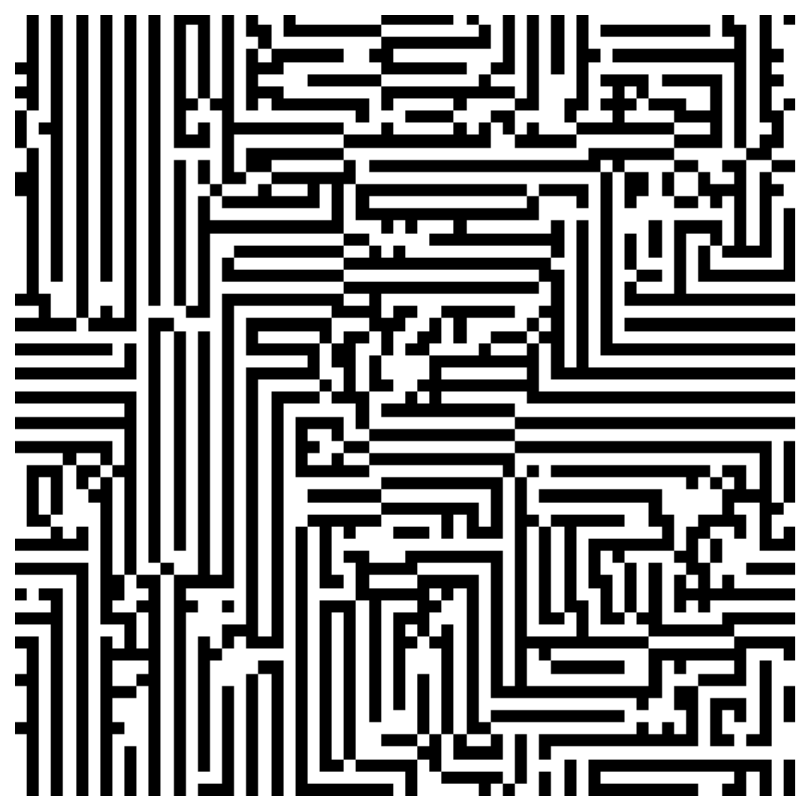}\label{fig:sa-rd-a-r-5}}
\subfigure[$r=6$]{\includegraphics[width=0.18\textwidth]{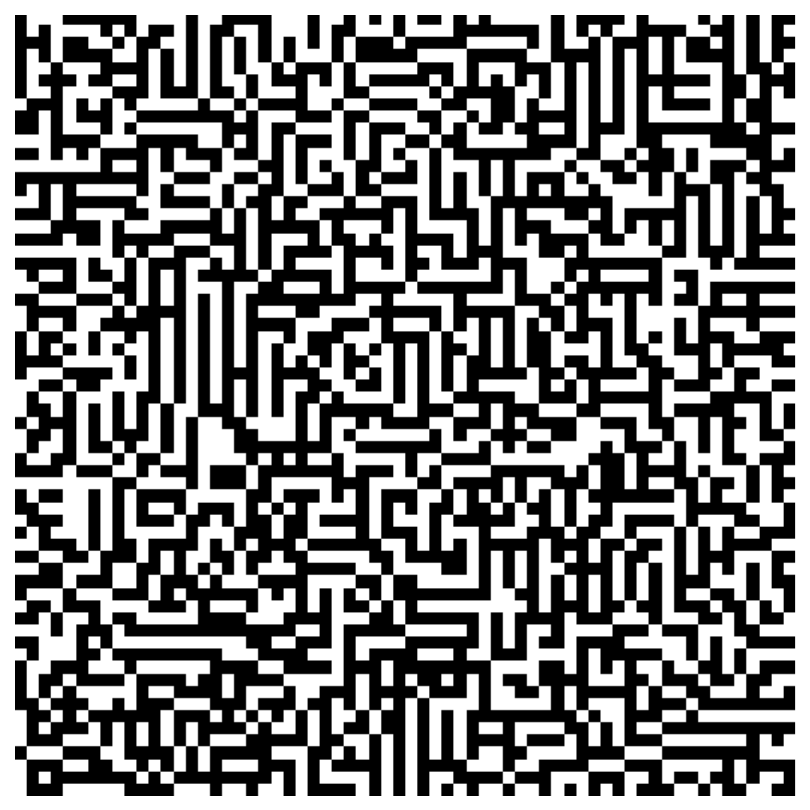}\label{fig:sa-rd-a-r-6}}
\subfigure[$r=7$]{\includegraphics[width=0.18\textwidth]{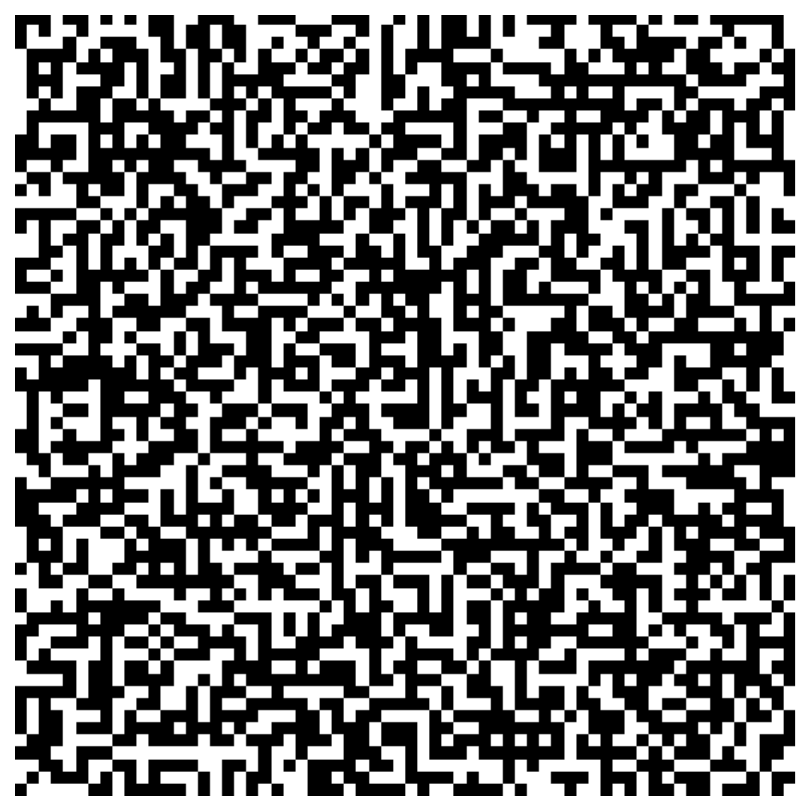}\label{fig:sa-rd-a-r-7}}
\subfigure[$r=8$]{\includegraphics[width=0.18\textwidth]{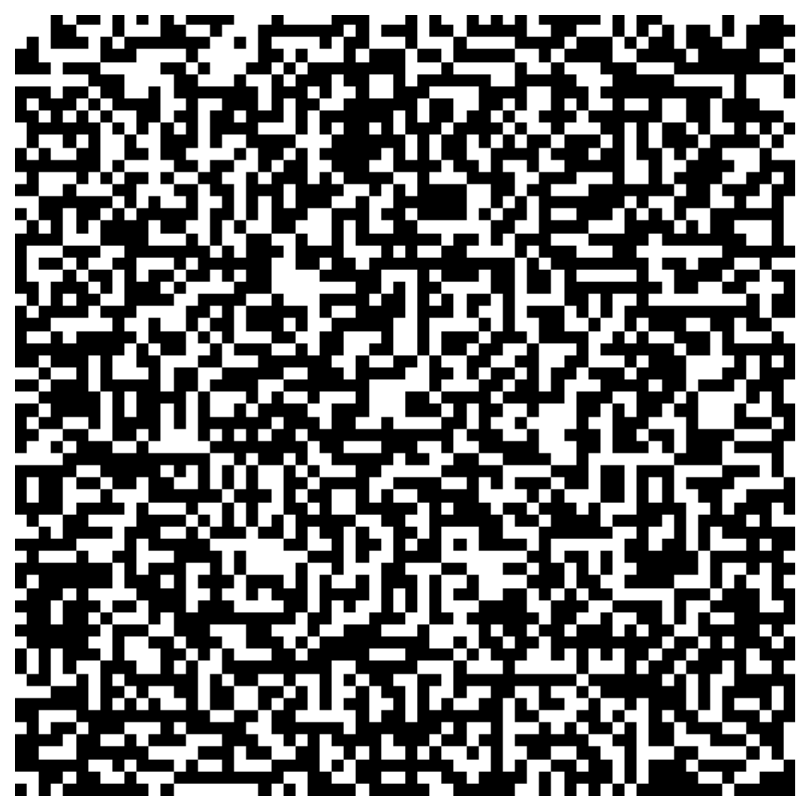}\label{fig:sa-rd-a-r-8}}

\caption{Typical configurations for deterministic and probabilistic mechanisms of rule switching. In the probabilistic case, the rules are selected with equal probability. Examples illustrate the final configurations after 250 steps of simulation for the second threshold, $r=4,5,6,7,8$, for the synchronous -- panels \subref{fig:sa-rd-s-dp2-4} to \subref{fig:sa-rd-s-dp2-8} and \subref{fig:sa-rd-s-r-4}  to \subref{fig:sa-rd-s-r-8} -- and the asynchronous -- panels \subref{fig:sa-rd-a-dp2-4} to \subref{fig:sa-rd-a-dp2-8} and \subref{fig:sa-rd-a-r-4} to \subref{fig:sa-rd-a-r-8}  -- version of the updating policy. In the deterministic case, the standard and the alternative rules are applied interchangeably. In the probabilistic case, the choice of the rules is made using a bit flip. From the obtained configurations, one can conclude that the random selection mechanism leads to a higher density of living cells (cf. panels \subref{fig:sa-rd-a-r-8} and \subref{fig:sa-rd-s-r-8} with panels \subref{fig:sa-rd-a-dp2-8} and \subref{fig:sa-rd-s-dp2-8}). It is also visible that the long-range patterns observed in the asynchronous case are more robust for the random rule selection mechanism (cf. panels \subref{fig:sa-rd-a-dp2-5} to \subref{fig:sa-rd-a-dp2-8} with panels \subref{fig:sa-rd-a-r-5} to \subref{fig:sa-rd-a-r-8}). This suggests that random selection of rules leads to a more stable behaviour of the system. It is also worth noting that the deterministic selection of rules does not increase the density of living cells for the synchronous updating policy (cf. panels \subref{fig:sa-rd-s-dp2-4} to \subref{fig:sa-rd-s-dp2-8}). Thus, the randomness of rule selection is essential for the observed effects.}
\label{fig:sa-rd}
\end{figure*}

Let us start by discussing the main differences one can observe in the final configurations with different parameters used by the rule selection mechanism. For this purpose, we describe the results of the numerical experiments based on the described procedures. Unless stated otherwise, in each case we consider a  lattice $L\times L$ with $L=2^6=64$, with periodic boundary conditions to minimize the finite-size effect. To ensure that the state of the lattice is representative, we use 250 simulation steps, each one consisting of playing the game by all agents. 

In the synchronous update policy, corresponding to the standard version of the GoL, the state of the lattice is updated globally at the end of each simulation step.
Thus, the state of the lattice is calculated using the state obtained in the previous step.

On the other hand, in the asynchronous policy, the state of each agent is calculated and updated immediately. Thus, each simulation step consists of the calculation of the next state of the cell and the update of the cell state. The order of cells is chosen randomly~\cite{comer2014who}. However, one should note that even for the deterministic choice of cells, the asynchronous and the synchronous policies will lead to different behaviours.

The first observation one can make from the comparison of synchronous versus asynchronous updating is the latter leads to the formation of stable, long-range patterns. From the examples presented in \Fref{fig:patern-formation}, one can see that the patters formed for the values of the threshold $r>4$ have homogeneity increasing with both the value of $r$ and the probability of utilizing the rule based on the higher threshold. This effect is not visible for the synchronous updating. Thus, one can conclude the asynchronous policy leads to more predictable pattern formation, while the synchronous policy leads to unstable patterns, even for alternative rules with high resistance for overpopulation.

It is also interesting to note the difference in the behaviour when comparing the deterministic and the probabilistic rule selection mechanisms for a fixed alternative threshold. In the probabilistic case, from the configurations presented in \Fref{fig:sa-rd} one can conclude that the random selection mechanism leads to a higher density of living cells (cf. \Fref{fig:sa-rd-a-dp2-8} and \Fref{fig:sa-rd-a-r-8} with  \Fref{fig:sa-rd-s-dp2-8} and \Fref{fig:sa-rd-a-dp2-8}). It is also visible that the long-range patterns observed in the asynchronous case are more robust for the random rule selection mechanism (cf. \Fref{fig:sa-rd-a-dp2-5} to \ref{fig:sa-rd-a-dp2-8} with \Fref{fig:sa-rd-a-r-5} to \ref{fig:sa-rd-a-r-8}). This suggests that the random selection of rules leads to a more stable behaviour of the system. Additionally, one can note that the deterministic selection of rules does not increase the density of living cells for the synchronous updating policy (cf.~\Fref{fig:sa-rd-s-dp2-4} to \ref{fig:sa-rd-s-dp2-8}). For this reason, one can see that the randomness of rule selection is essential for the observed effects, and it is preferred over the deterministic rule switching mechanism.

\subsection{Stability}

Another aspect of the considered model is the stability of the pattern formation. From the results presented  in \Fref{fig:patern-formation} and \Fref{fig:sa-rd} one can observe the final configurations of the system. However, it is well-known that in the standard GoL, the final configurations is rarely static.

\begin{figure*}[ht!]
	\centering
	\subfigure[$r=6$]{\label{fig:main-st6-il20}
		\includegraphics[scale=0.9]{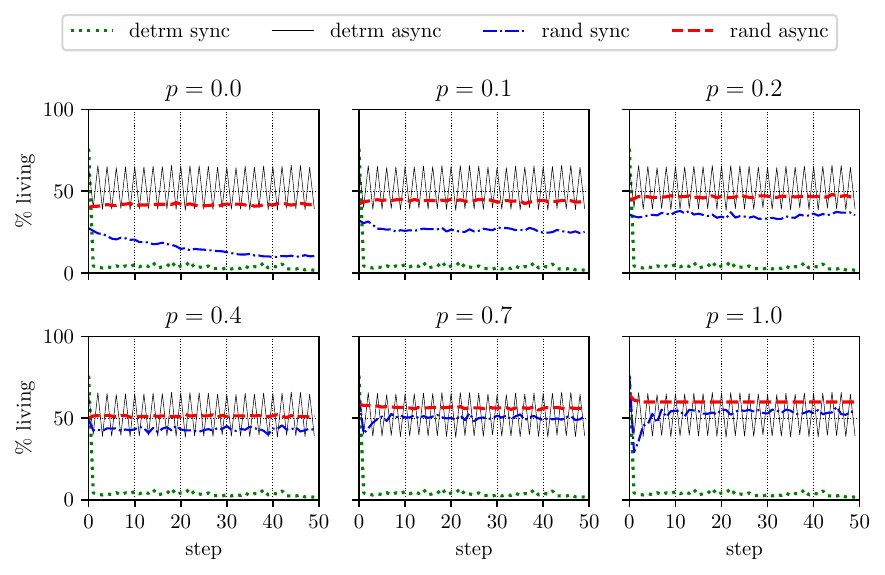}}
	\subfigure[$r=8$]{\label{fig:main-st8-il20}
		\includegraphics[scale=0.9]{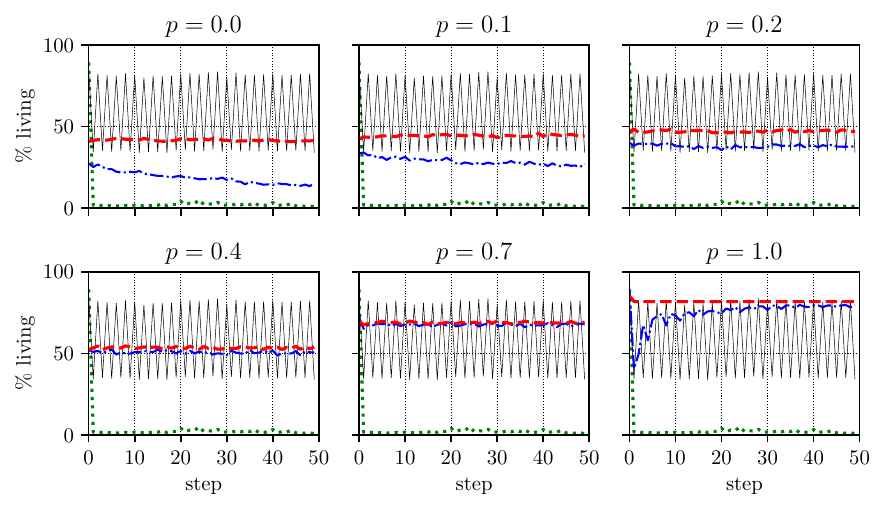}}
	\caption{Sample realization for the percentage of living cells for the GoL with different rule switching mechanisms and for synchronous and asynchronous cases. In this case, the games were played with the second threshold set to \subref{fig:main-st6-il20} $r=6$ and \subref{fig:main-st8-il20} $r=8$. For the random rule switching mechanics, each rule is applied with probability $\frac{1}{2}$. One should note that for the deterministic case, rules are switched after each step. For both cases, both synchronization policies were used. Each plot contains realizations of four possible scenarios. For the higher values of the alternative threshold, the system can achieve higher number of the living cells. However, this is possible only for the asynchronous updating policy. Random rule selection mechanism is more effective in both synchronous and asynchronous scenarios. Additionally, for the deterministic rule selection mechanisms, it is impossible to achieve a stable number of living cells.}
	\label{fig:realizations}
\end{figure*}

In this case, the most interesting observation is the interplay between the synchronization and the stability. From the results presented in \Fref{fig:patern-formation}, one can observe that the system can achieve higher number of the living cells with the increasing value of the threshold used in the alternative rules. This is expected, as the increase in the value of the alternative threshold might be understood as a immunity against overpopulation.

However, as one can see in \Fref{fig:realizations}, this effect is more prominent for the asynchronous updating policy. Moreover, in random rule selection mechanism it more effective in both synchronous and asynchronous scenario. In the synchronous policy, the growth in the number of living cells is slower than in the  asynchronous case. This suggests that asynchronous policy makes the system more suitable for growth. Additionally, for the deterministic rule selection mechanisms, it is impossible to achieve a stable number of living cells. Thus, in the case of synchronous policy, deterministic rule selection leads to unstable behaviour. The above observations suggest that the random rule selection policy leads to more stable and more predictable growth in the system. 
 
\subsection{Synchronization and growth}

Finally, the main conclusion from the numerical results presented in the previous sections is that the introduced model clearly displays the trade-off between the stability and the growth. To assess the impact of the rule selection mechanism and the updating policy on the growth in the GoL, we calculate average number of living cells for all introduced scenarios and compare this vale with different values of the second threshold. 
\begin{figure*}[ht!]
	\centering
	\includegraphics[scale=1]{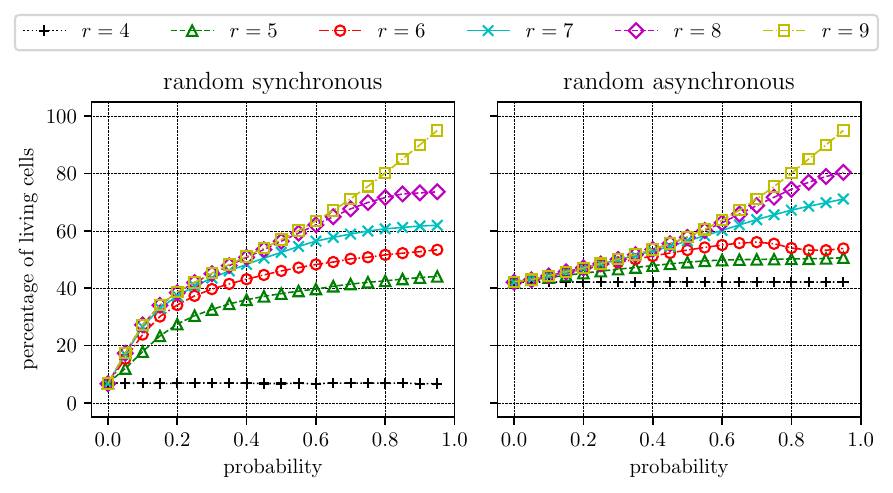}
	\caption{Average number of living cells for different values of probability in the random rule selection mechanism. Plotted results were obtained for synchronous (left) versus asynchronous (right) updating policy for different values of the second threshold. Average number of living cells for $64\times 64$ lattice, 250 steps, averaged over 200 realizations, for the values of the second threshold $r=4,5,\dots,9$. Data marked with black pluses depict the baseline case of the standard GoL. Data marked with yellow squares  depict the case $r=9$, representing the situation with the total immunity to overpopulation. One can see that in both cases for the second threshold $r=4$ and $r=5$ it is impossible to achieve above 50\% of the living cells. In the asynchronous case for $r=6$, the increase in the probability above $p\approx0.7$ leads to the decrease in the growth. For $r=7$ and $r=8$ one can observe that the asynchronous  updating policy leads to a stable growth in the system. Note that for the case with the second threshold $r=4$, the values do not depend on the probability.}
	\label{fig:mean-living}
\end{figure*}

The results for the random rule selection mechanisms are presented in \Fref{fig:mean-living}. In this situation, it is possible to distinguish two cases. For the second threshold $r=4$ and $r=5$, representing a small increase in the immunity comparing with the standard GoL, it is impossible to achieve more than 50\% of the living cells. Comparing this with the baseline of $\approx50\%$ in the asynchronous scenario, one can conclude that the small changes in the immunity are irrelevant in the case of the asynchronous updating. Additionally, in the asynchronous case for $r=6$, the increase in the probability above $p\approx0.7$ leads to the decrease in the growth. This can be explained considering that the increased immunity for overpopulation is in this case compensated by the increased growth. For $r=7$ and $r=8$ one can observe that for both updating policies such increase in the immunity is beneficial for the growth. It is also visible that the asynchronous updating policy leads to the stronger growth in the system. Moreover, one should note that the difference between the immunity levels is more prominent in the case of the synchronous policy, especially for the larger values of $p$. This leads to a conclusion that the asynchronous updating promotes the growth in the system, even for the case when a small fraction of population has increased immunity to overpopulation.

\subsection{Pattern complexity}

Let us now provide an attempt at grasping the quantitative difference in the spatial correlations and the form of structures observed during the evolution of the introduced models. To this end we adapt the approach utilized in \cite{pena2021life}, based on the analysis of the complexity of the patterns~\cite{bates1993measuring,wackerbauer1994comparative,andrienko2000complexity} formed by the GoL.

In this approach, the complexity of 2D patterns is described by the conditional entropy
\begin{equation}
H(x|y) = - \sum_{x,y} P(x,y) \log_2 P(x|y),
\end{equation}
where $P(x,y)$ is the joint probability of observing state $(x,y)$ of points $x$ and $y$, and $P(x|y)$ is the conditional probability of observing $x$ given $y$. Conditional entropy quantifies the amount of information required to describe the state of variable $x$, provided that the information about the state of the variable~$y$ is available. Hence, it can be understood as the amount of information required to describe the complexity.

In the case of 2D pattern, $x$ and $y$ represent discrete states of the cells, $x,y\in\{0,1\}$. For a given cell $x$, the conditional entropy is calculated as the average of the conditional entropies regarding its neighbourhood. For the purpose of our analysis we follow the approach from \cite{pena2021life} and we use the von Neumann neighbourhood. Hence, the complexity of the pattern is captured by averaging conditional entropies calculated in four directions,
\begin{equation}
C(x) = \frac{1}{4}\left[H(x|x+(1,0)) +  H(x|x+(-1,0)) + H(x|x+(0,1)) + H(x|x+(0,-1)) \right],
\end{equation}
where $x+(dx,dy)$ represents the cell at position with coordinates shifted by $(dx,dy)$. The conditional entropy of the full configuration is calculated as the average conditional entropy of all cells,
\begin{equation}
\tilde{C} = \frac{1}{|X|}\sum_{x\in X} C(x),
\end{equation}
where $X$ is the set of all cells.
One should note that the increasing value of $\tilde{C}$ corresponds to the increasing randomness of the configurations resulting from the rule applications. 

\begin{figure*}[t!]
	\centering
\includegraphics{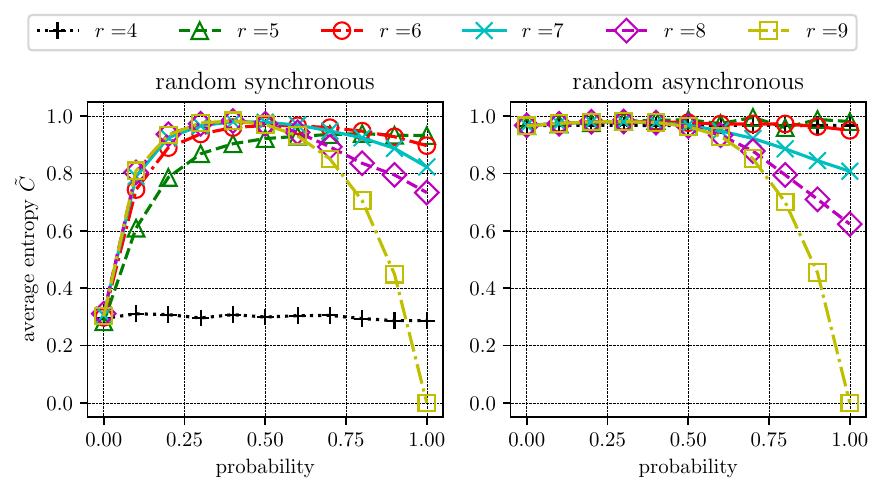}
	\caption{Complexity of the patterns generated for the synchronous and asynchronous scenarios for the GoL for different values of random rule switching probabilities. Each line represents a variant with the game with a specified second threshold. The data were obtained for averaging over $10^3$ realizations of the process, with each process evolving for 250 steps. In this case the lattice size was set to $32\times 32$. Note that for the case with the second threshold $r=4$, the values do not depend on the probability.}
	\label{fig:pattern-mutent}
\end{figure*}

The results for the average mutual entropy for the random rules with different second threshold are provided in Figure~\ref{fig:pattern-mutent}. One can observe that both for the synchronous and for the asynchronous updating regimes, one can reduce the entropy by increasing the probability of utilizing the alternative rule. In the case of the synchronous updating, the entropy of the generated patterns increases with the probability of utilizing the alternative rules, as long as $p<0.5$. For the values of $p>0.5$, the effects of entropy reduction occur in both the synchronous and the asynchronous case for all values of the alternative threshold $r>5$. However, for values $p>0.5$ the effect is more noticeable in the case of the asynchronous regime for $r\geq7$. Interestingly, for the case of $r=9$, corresponding to introducing total immunity to overpopulation, for $p<0.5$, the effect of entropy increase is observed in the synchronous regime. On the other hand, for $r>0.5$, for both regimes, with the increasing domination of the rule promoting total immunity to overpopulation, the mutual entropy is decreasing similarly.

\section{Final remarks}\label{sec:final}

In the presented work we introduced an extended version of the GoL cellular automata incorporating randomness into the rule selection policy. We also provided an insight into the connections between the synchronization policy and the rule selection in the GoL, and their impact on the dynamics of the system. The introduced version of the GoL is extended with the possibility of selecting one of the rules during the evolution. Thanks to this, we relaxed the condition of using only a single rule during the evolution. As such, the introduced model can be used to modify a policy for rules selection during the evolution by modifying the probability of rule selection depending on the state of the system. Hence, the proposed version can be more flexible in applications related to biological and social systems.

One should note that the presented models significantly extend the models of 1D stochastic GoL developed in \cite{monetti1997stochastic}, where the randomness of the model was due to the presence of noise. In the GoL with random rules, the randomness is controlled in the sense that it can be tuned to provide the desired effect. Hence, the presented model is richer and provides a flexible tool for describing complex systems.

For the purpose of the presented study we used only simple mechanisms -- deterministic, where the rules are exchanged after each step, or random, where the rule is chosen with some fixed probability. In both cases, we do not take into account the memory. We also do not adapt the rule selection to the current state of the cell or neighbouring cells. Such mechanism could be included by altering the probability of choosing one of the rules depending on the population of the living cells in the neighbourhood.

Moreover, we have limited our investigation to the case where the rule switching mechanisms do not depend on the history. However, one can consider the version of the rule switching mechanism where the selection of a rule at the current time-step depends on the history of the cell. In particular, it is possible to consider a mechanism in which the cell tends to decrease its resistance to the overpopulation. This is equivalent to increasing the probability of using an alternative rule growing with the living time of the cell. Such a mechanism would be useful for incorporating into the GoL a kind of acquired immunity. As the cellular automata stores the memory of the previous moves in its states, it is natural to use the current state of the cell to choose the rule to be applied at a given step. These mechanisms are actually applied in the standard case of the Game of Life, where the next state is calculated using the number of living cells in the Moore neighbourhood and using the current state. However, by introducing the additional parameter stored by each cell, one can significantly increase the complexity of the studied model.

Another method for studying the effects of the variable rule mechanism on the formed patterns could be based on the underlying connectivity of the resulting graphs. However, in this case one has to take into account that the formed patterns are not static, and the dynamics has to be included into the process of computing metrics such as clustering coefficient.

The presented work provides a simple example of the modification of the GoL which can lead to interesting new behaviour. Recent applications of probabilistic cellular automata in physics~\cite{wetterich2022fermionic} and chemistry \cite{yanez2019automaton} provide some of the examples of using such automata. This suggests that by combining the asynchronous updating policy with the freedom of choosing the updating rules, one can describe many interesting models, capable of grasping the rich behaviour patterns in complex physical and biological systems.

\section*{Acknowledgments}

Author would like to thank Nenggang Xie, Ye Ye and Wang Meng for motivating the presented study and for their hospitality during the visit at the Anhui University of Technology, Ludmila Botelho for pointing references related to the quantum Game of Life, Krzysztof Domino for interesting discussions concerning agent-based modelling and data analysis, Izabela Miszczak for proofreading the manuscript, and anonymous reviewers for motivating reviews.

This work has been motivated by personal curiosity and received no funding from any agency. 

\appendix

\section{Model implementation}

This model used to execute numerical experiments is implemented in NetLogo language as a version of GoL cellular 2D automaton extended with the ability to alter the rules utilized by the cell during the evolution. The implementation of the model can be downloaded from~\cite{gol-mc}. Additionally, in order to facilitate the reproducibility, the model, description of experiments, controlling scripts, and visualization scripts were made available from public repository~\cite{gol-github}.

Simple versions of random and deterministic mechanisms of rules selection are implemented. In both case, at each step, the cell can be updated according to one of the rules - the standard one and the alternative one. The standard rule is identical to the rule for dying due to overpopulation (Any alive cell with four or more alive neighbours dies, because of the  overpopulation). The alternative rule for threshold $r$ is: Any alive cell with $r$ of more alive neighbours dies, because of the  overpopulation.

Additionally, one can switch between synchronous and asynchronous state updating policies. In the synchronous updating policy, corresponding to the standard version of GoL, the state of the lattice is updated globally at the end of each simulation step. In the asynchronous policy, the state of each agent is calculated and updated immediately. Thus, each simulation step consists of the calculation of the next state of the cell, and the update of the cell state. The cells are updated in random order randomly.

\begin{figure}[h]
\includegraphics[width=0.9\textwidth]{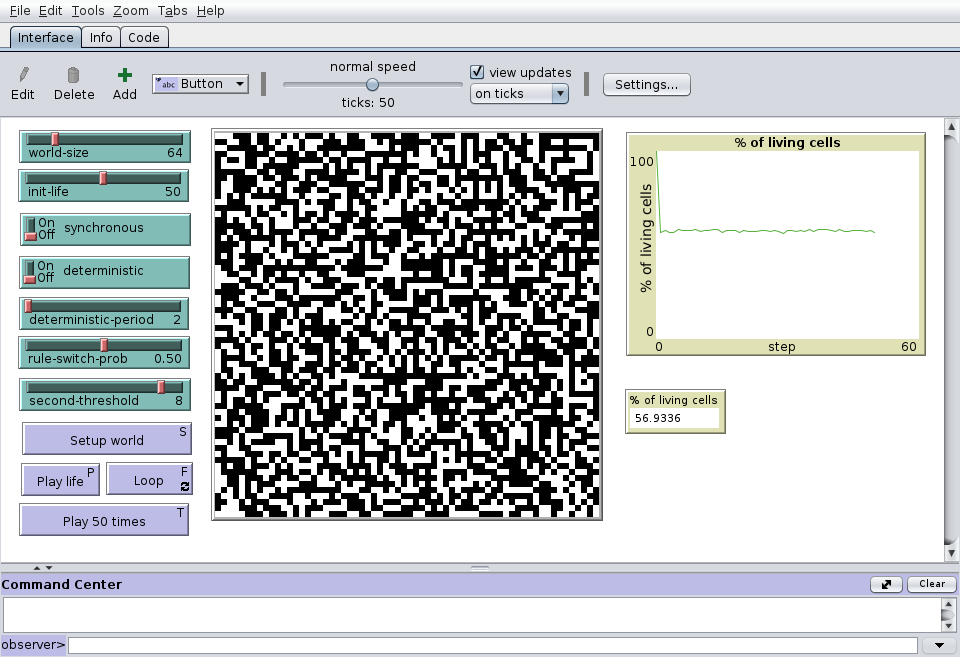}
\caption{User interface of the NetLogo with loaded model of Game of Life with rule switching~\cite{gol-mc}.}
\label{fig:life-rule-switching-gui}
\end{figure}

The following parameters of the model are available through the controls (see \Fref{fig:life-rule-switching-gui}):

\begin{itemize}
\item \texttt{world-size} - size of the lattice used to run the simulation;
\item \texttt{init-life} - percentage of living cells at the beginning of the simulation;
\item \texttt{synchronous} - toggle between synchronous and asynchronous updating policy;
\item \texttt{deterministic} - toggle between deterministic and random rule selection mechanism;
\item \texttt{deterministic-period} - the number of iterations between the utilizations of the alternative rule in the deterministic rule switching mechanism;
\item \texttt{rule-switch-prob} - set the probability of utilizing an alternative rule in the random rule switching mechanism;
\item \texttt{second-threshold} - threshold used in the second (alternative) rule used in the game.
\end{itemize}

After setting the required parameters, use \texttt{Setup world} button to initialize the simulation. To run the model ones, use \texttt{Play life} button. \texttt{Loop} button runs the game in a loop, and \texttt{Play 50 times} runs 50 simulation cycles.

\section*{References}

\bibliographystyle{elsarticle-num}
\bibliography{gol_rule_switching}

\begin{thebibliography}{10}
\expandafter\ifx\csname url\endcsname\relax
  \def\url#1{\texttt{#1}}\fi
\expandafter\ifx\csname urlprefix\endcsname\relax\def\urlprefix{URL }\fi
\expandafter\ifx\csname href\endcsname\relax
  \def\href#1#2{#2} \def\path#1{#1}\fi

\bibitem{vonneumann1966theory}
J.~Von~Neumann,
  \href{https://archive.org/details/theoryofselfrepr00vonn}{Theory of
  self-reproducing automata}, University of Illinois Press, 1966.
\newline\urlprefix\url{https://archive.org/details/theoryofselfrepr00vonn}

\bibitem{schiff2011cellular}
J.~L. Schiff, Cellular automata: a discrete view of the world, John Wiley \&
  Sons, 2011.
\newblock \href {https://doi.org/10.1002/9781118032381}
  {\path{doi:10.1002/9781118032381}}.

\bibitem{chopard1998cellular}
B.~Chopard, M.~Droz, Cellular Automata Modeling of Physical Systems, Cambridge
  University Press, 1998.
\newblock \href {https://doi.org/10.1017/CBO9780511549755}
  {\path{doi:10.1017/CBO9780511549755}}.

\bibitem{hooft2016cellular}
G.~t~Hooft, The Cellular Automaton Interpretation of Quantum Mechanics,
  Springer-Verlag GmbH, 2016.
\newblock \href {http://arxiv.org/abs/1405.1548v3} {\path{arXiv:1405.1548v3}},
  \href {https://doi.org/10.1007/978-3-319-41285-6}
  {\path{doi:10.1007/978-3-319-41285-6}}.

\bibitem{wolfram1994cellulara}
S.~Wolfram, Cellular automata and complexity : collected papers, Addison-Wesley
  Pub. Co, Reading, Mass, 1994.

\bibitem{wolfram1984cellular}
S.~Wolfram, Cellular automata as models of complexity, Nature 311~(5985) (1984)
  419--424.
\newblock \href {https://doi.org/10.1038/311419a0}
  {\path{doi:10.1038/311419a0}}.

\bibitem{rendell2011universal}
P.~Rendell, A universal {Turing} machine in {Conway{\textquotesingle}s Game of
  Life}, in: 2011 International Conference on High Performance Computing \&
  Simulation, {IEEE}, 2011.
\newblock \href {https://doi.org/10.1109/hpcsim.2011.5999906}
  {\path{doi:10.1109/hpcsim.2011.5999906}}.

\bibitem{kier1999cellular}
L.~B. Kier, C.-K. Cheng, B.~Testa, Cellular automata models of biochemical
  phenomena, Future Generation Computer Systems 16~(2-3) (1999) 273--289.
\newblock \href {https://doi.org/10.1016/s0167-739x(99)00052-7}
  {\path{doi:10.1016/s0167-739x(99)00052-7}}.

\bibitem{korte2013cellular}
A.~de~Korte, H.~Brouwers, A cellular automata approach to chemical reactions; 1
  reaction controlled systems, Chemical Engineering Journal 228 (2013)
  172--178.
\newblock \href {https://doi.org/10.1016/j.cej.2013.04.084}
  {\path{doi:10.1016/j.cej.2013.04.084}}.

\bibitem{pca2018}
P.-Y. Louis, F.~R. Nardi (Eds.), Probabilistic Cellular Automata, Vol.~27,
  Springer International Publishing, 2018.
\newblock \href {https://doi.org/10.1007/978-3-319-65558-1}
  {\path{doi:10.1007/978-3-319-65558-1}}.

\bibitem{bhattacharjee2018survey}
K.~Bhattacharjee, N.~Naskar, S.~Roy, S.~Das, A survey of cellular automata:
  types, dynamics, non-uniformity and applications, Natural Computing 19~(2)
  (2018) 433--461.
\newblock \href {https://doi.org/10.1007/s11047-018-9696-8}
  {\path{doi:10.1007/s11047-018-9696-8}}.

\bibitem{fates2014guided}
N.~Fatès, A guided tour of asynchronous cellular automata, Journal of Cellular
  Automata (2014).
\newblock \href {https://doi.org/10.1007/978-3-642-40867-0_2}
  {\path{doi:10.1007/978-3-642-40867-0_2}}.

\bibitem{baetens2012effect}
J.~Baetens, P.~V. der Weeën, B.~D. Baets, Effect of asynchronous updating on
  the stability of cellular automata, Chaos, Solitons {\&} Fractals 45~(4)
  (2012) 383--394.
\newblock \href {https://doi.org/10.1016/j.chaos.2012.01.002}
  {\path{doi:10.1016/j.chaos.2012.01.002}}.

\bibitem{reia2015nonsynchronous}
S.~M. Reia, O.~Kinouchi, Nonsynchronous updating in the multiverse of cellular
  automata, Physical Review E 91~(4) (2015) 042110.
\newblock \href {https://doi.org/10.1103/physreve.91.042110}
  {\path{doi:10.1103/physreve.91.042110}}.

\bibitem{boure2012probing}
O.~Bour{\'{e}}, N.~Fat{\`{e}}s, V.~Chevrier, Probing robustness of cellular
  automata through variations of asynchronous updating, Natural Computing
  11~(4) (2012) 553--564.
\newblock \href {https://doi.org/10.1007/s11047-012-9340-y}
  {\path{doi:10.1007/s11047-012-9340-y}}.

\bibitem{boure2011robustness}
O.~Bour{\'{e}}, N.~Fat{\`{e}}s, V.~Chevrier, Robustness of cellular automata in
  the light of asynchronous information transmission, in: Lecture Notes in
  Computer Science, Springer Berlin Heidelberg, 2011, pp. 52--63.
\newblock \href {https://doi.org/10.1007/978-3-642-21341-0_11}
  {\path{doi:10.1007/978-3-642-21341-0_11}}.

\bibitem{berlekamp2001winning}
E.~R. Berlekamp, J.~H. Conway, R.~K. Guy, Winning Ways for Your Mathematical
  Plays, Volume 1, AK Peters, Ltd., 2001.
\newblock \href {https://doi.org/10.1007/s00283-021-10097-3}
  {\path{doi:10.1007/s00283-021-10097-3}}.

\bibitem{peper2010variations}
F.~Peper, S.~Adachi, J.~Lee, Variations on the game of life, in: Game of Life
  Cellular Automata, Springer London, 2010, pp. 235--255.
\newblock \href {https://doi.org/10.1007/978-1-84996-217-9_13}
  {\path{doi:10.1007/978-1-84996-217-9_13}}.

\bibitem{blok1999synchronous}
H.~J. Blok, B.~Bergersen, Synchronous versus asynchronous updating in the
  ``game of life'', Phys. Rev. E 59 (1999) 3876--3879.
\newblock \href {https://doi.org/10.1103/PhysRevE.59.3876}
  {\path{doi:10.1103/PhysRevE.59.3876}}.

\bibitem{lee2004asynchronous}
J.~Lee, S.~Adachi, F.~Peper, K.~Morita, Asynchronous game of life, Physica D:
  Nonlinear Phenomena 194~(3-4) (2004) 369--384.
\newblock \href {https://doi.org/10.1016/j.physd.2004.03.007}
  {\path{doi:10.1016/j.physd.2004.03.007}}.

\bibitem{poindron2021general}
A.~Poindron, A general model of binary opinions updating, Mathematical Social
  Sciences 109 (2021) 52--76.
\newblock \href {https://doi.org/10.1016/j.mathsocsci.2020.10.004}
  {\path{doi:10.1016/j.mathsocsci.2020.10.004}}.

\bibitem{billings2003indentification}
S.~Billings, Y.~Yang, Identification of probabilistic cellular automata, IEEE
  Transactions on Systems, Man, and Cybernetics, Part B (Cybernetics) 33~(2)
  (2003) 225--236.
\newblock \href {https://doi.org/10.1109/TSMCB.2003.810437}
  {\path{doi:10.1109/TSMCB.2003.810437}}.

\bibitem{agapie2014probabilisti}
A.~Agapie, A.~Andreica, M.~Giuclea, Probabilistic cellular automata, Journal of
  Computational Biology 21~(9) (2014) 699--708, pMID: 24999557.
\newblock \href {https://doi.org/10.1089/cmb.2014.0074}
  {\path{doi:10.1089/cmb.2014.0074}}.

\bibitem{mairesse2014probabilistic}
J.~Mairesse, I.~Marcovici, Around probabilistic cellular automata, Theoretical
  Computer Science 559 (2014) 42--72.
\newblock \href {https://doi.org/10.1016/j.tcs.2014.09.009}
  {\path{doi:10.1016/j.tcs.2014.09.009}}.

\bibitem{gravner2021periodic}
J.~Gravner, X.~Liu, Periodic solutions of one-dimensional cellular automata
  with uniformly chosen random rules, The Electronic Journal of Combinatorics
  28 (2021) P4.51--P4.51.
\newblock \href {https://doi.org/10.37236/10114} {\path{doi:10.37236/10114}}.

\bibitem{gravner2022one}
J.~Gravner, X.~Liu, One-dimensional cellular automata with random rules:
  longest temporal period of a periodic solution, Electronic Journal of
  Probability 27~(none) (2022).
\newblock \href {https://doi.org/10.1214/22-ejp744}
  {\path{doi:10.1214/22-ejp744}}.

\bibitem{aguileravenegas2019probabilistic}
G.~Aguilera-Venegas, J.~L. Gal{\'{a}}n-Garc{\'{\i}}a, R.~Egea-Guerrero,
  M.~{\'{A}}. Gal{\'{a}}n-Garc{\'{\i}}a, P.~Rodr{\'{\i}}guez-Cielos,
  Y.~Padilla-Dom{\'{\i}}nguez, M.~Gal{\'{a}}n-Luque, A probabilistic extension
  to {Conway}'s {Game} {of} {Life}, Advances in Computational Mathematics
  45~(4) (2019) 2111--2121.
\newblock \href {https://doi.org/10.1007/s10444-019-09696-8}
  {\path{doi:10.1007/s10444-019-09696-8}}.

\bibitem{gabor2021probabilistic}
T.~Gabor, M.~L. Rosenfeld, C.~Linnhoff-Popien, {A Probabilistic Game of Life on
  a Quantum Annealer}, Vol. ALIFE 2021: The 2021 Conference on Artificial Life
  of ALIFE 2021: The 2021 Conference on Artificial Life, 2021, 103.
\newblock \href {https://doi.org/10.1162/isal_a_00441}
  {\path{doi:10.1162/isal_a_00441}}.

\bibitem{mullick2019effect}
P.~Mullick, P.~Sen, Effect of bias in a reaction-diffusion system in two
  dimensions, Physical Review E 99~(5) (2019) 052123.
\newblock \href {https://doi.org/10.1103/physreve.99.052123}
  {\path{doi:10.1103/physreve.99.052123}}.

\bibitem{pavlic2014self}
T.~Pavlic, A.~Adams, P.~Davies, S.~Walker, Self-referencing cellular automata:
  A model of the evolution of information control in biological systems, in:
  Artificial Life 14: Proceedings of the Fourteenth International Conference on
  the Synthesis and Simulation of Living Systems, The {MIT} Press, 2014.
\newblock \href {https://doi.org/10.7551/978-0-262-32621-6-ch083}
  {\path{doi:10.7551/978-0-262-32621-6-ch083}}.

\bibitem{chan2019lenia}
B.~W.-C. Chan, Lenia: Biology of artificial life, Complex Systems 28~(3) (2019)
  251--286.
\newblock \href {https://doi.org/10.25088/complexsystems.28.3.251}
  {\path{doi:10.25088/complexsystems.28.3.251}}.

\bibitem{chan2020lenia}
B.~W.-C. Chan, Lenia and expanded universe, in: The 2020 Conference on
  Artificial Life, {MIT} Press, 2020.
\newblock \href {https://doi.org/10.1162/isal_a_00297}
  {\path{doi:10.1162/isal_a_00297}}.

\bibitem{miszczak2012high-level}
J.~A. Miszczak, High-level structures for quantum computing, Vol.~4, Morgan \&
  Claypool Publishers, 2012.
\newblock \href {https://doi.org/10.2200/S00422ED1V01Y201205QMC006}
  {\path{doi:10.2200/S00422ED1V01Y201205QMC006}}.

\bibitem{arrighi2010quantum}
P.~Arrighi, J.~Grattage, A quantum game of life, in: J.~Kari (Ed.), Second
  Symposium on Cellular Automata "Journ{\'{e}}es Automates Cellulaires", {JAC}
  2010, Turku, Finland, December 15-17, 2010. Proceedings, Turku Center for
  Computer Science, 2010, pp. 31--42.
\newblock \href {https://doi.org/10.48550/arXiv.1010.3120}
  {\path{doi:10.48550/arXiv.1010.3120}}.

\bibitem{bleh2012quantum}
D.~Bleh, T.~Calarco, S.~Montangero, Quantum game of life, EPL (Europhysics
  Letters) 97~(2) (2012) 20012.
\newblock \href {https://doi.org/10.1209/0295-5075/97/20012}
  {\path{doi:10.1209/0295-5075/97/20012}}.

\bibitem{ney2022entanglement}
P.-M. Ney, S.~Notarnicola, S.~Montangero, G.~Morigi, Entanglement in the
  quantum game of life, Phys. Rev. A 105 (2022) 012416.
\newblock \href {http://arxiv.org/abs/2104.14924} {\path{arXiv:2104.14924}},
  \href {https://doi.org/10.1103/PhysRevA.105.012416}
  {\path{doi:10.1103/PhysRevA.105.012416}}.

\bibitem{hillberry2021entangled}
L.~E. Hillberry, M.~T. Jones, D.~L. Vargas, P.~Rall, N.~Y. Halpern, N.~Bao,
  S.~Notarnicola, S.~Montangero, L.~D. Carr, Entangled quantum cellular
  automata, physical complexity, and goldilocks rules, Quantum Science and
  Technology 6~(4) (2021) 045017.
\newblock \href {https://doi.org/10.1088/2058-9565/ac1c41}
  {\path{doi:10.1088/2058-9565/ac1c41}}.

\bibitem{jones2022small}
E.~B. Jones, L.~E. Hillberry, M.~T. Jones, M.~Fasihi, P.~Roushan, Z.~Jiang,
  A.~Ho, C.~Neill, E.~Ostby, P.~Graf, E.~Kapit, L.~D. Carr, Small-world complex
  network generation on a digital quantum processor, Nature Communications
  13~(1) (aug 2022).
\newblock \href {https://doi.org/10.1038/s41467-022-32056-y}
  {\path{doi:10.1038/s41467-022-32056-y}}.

\bibitem{monetti1997stochastic}
R.~A. Monetti, E.~V. Albano, Stochastic game of life in one dimension, Physica
  A: Statistical Mechanics and its Applications 234~(3) (1997) 785--791.
\newblock \href {https://doi.org/10.1016/S0378-4371(96)00316-0}
  {\path{doi:10.1016/S0378-4371(96)00316-0}}.

\bibitem{comer2014who}
K.~W. Comer, \href{http://mars.gmu.edu/handle/1920/9070}{{Who Goes First? An
  Examination of the Impact of Activation on Outcome Behavior in Agent-based
  Models}}, Ph.D. thesis (2014).
\newline\urlprefix\url{http://mars.gmu.edu/handle/1920/9070}

\bibitem{pena2021life}
E.~Pe{\~{n}}a, H.~Sayama, Life worth mentioning: Complexity in life-like
  cellular automata, Artificial Life 27~(2) (2021) 105--112.
\newblock \href {https://doi.org/10.1162/artl_a_00348}
  {\path{doi:10.1162/artl_a_00348}}.

\bibitem{bates1993measuring}
J.~E. Bates, H.~K. Shepard, Measuring complexity using information fluctuation,
  Physics Letters A 172~(6) (1993) 416--425.
\newblock \href {https://doi.org/10.1016/0375-9601(93)90232-o}
  {\path{doi:10.1016/0375-9601(93)90232-o}}.

\bibitem{wackerbauer1994comparative}
R.~Wackerbauer, A.~Witt, H.~Atmanspacher, J.~Kurths, H.~Scheingraber, A
  comparative classification of complexity measures, Chaos, Solitons \&
  Fractals 4~(1) (1994) 133--173.
\newblock \href {https://doi.org/10.1016/0960-0779(94)90023-x}
  {\path{doi:10.1016/0960-0779(94)90023-x}}.

\bibitem{andrienko2000complexity}
Y.~Andrienko, N.~Brilliantov, J.~Kurths, Complexity of two-dimensional
  patterns, The European Physical Journal B 15~(3) (2000) 539--546.
\newblock \href {https://doi.org/10.1007/s100510051157}
  {\path{doi:10.1007/s100510051157}}.

\bibitem{wetterich2022fermionic}
C.~Wetterich, Fermionic quantum field theories as probabilistic cellular
  automata, Phys. Rev. D 105 (2022) 074502.
\newblock \href {https://doi.org/10.1103/PhysRevD.105.074502}
  {\path{doi:10.1103/PhysRevD.105.074502}}.

\bibitem{yanez2019automaton}
O.~Yañez, R.~Báez-Grez, D.~Inostroza, W.~A. Rabanal-León, R.~Pino-Rios,
  J.~Garza, W.~Tiznado, {AUTOMATON}: A program that combines a probabilistic
  cellular automata and a genetic algorithm for global minimum search of
  clusters and molecules, Journal of Chemical Theory and Computation 15 (2019)
  1463--1475.
\newblock \href {https://doi.org/10.1021/acs.jctc.8b00772}
  {\path{doi:10.1021/acs.jctc.8b00772}}.

\bibitem{gol-mc}
J.~Miszczak, \href{http://modelingcommons.org/browse/one_model/7089}{{Conway's
  Game of Life} with rule switching} (2022).
\newline\urlprefix\url{http://modelingcommons.org/browse/one_model/7089}

\bibitem{gol-github}
J.~A. Miszczak,
  \href{https://github.com/jmiszczak/gol_rule_switching}{Implementation of the
  game of life with rule switching -- the model, description of experiments,
  controlling scripts, and visualization scripts} (2023).
\newblock \href {https://doi.org/10.5281/zenodo.8099606}
  {\path{doi:10.5281/zenodo.8099606}}.
\newline\urlprefix\url{https://github.com/jmiszczak/gol_rule_switching}

\end{thebibliography}

\end{document}